\documentclass[useAMS,usenatbib]{mn2e}

\usepackage{graphicx}

\def\hst{{\it HST}}
\def\H0{{\rm ~km~s^{-1}~Mpc^{-1}}}

\title[Halo star clusters of M31]{The structure of star clusters in the outer halo of M31}
\author[N. R. Tanvir et al.]
{N.~R.~Tanvir$^{1}$\thanks{E-mail: nrt3@star.le.ac.uk}, 
A.~D.~Mackey$^{2,3}$, A.~M.~N.~Ferguson$^{3}$, A.~Huxor$^{4}$, 
\newauthor
J.~I.~Read$^{5,1}$, G.~F.~Lewis$^6$,  M. J. Irwin$^7$, S. Chapman$^7$, R. Ibata$^8$, 
\newauthor
M. I. Wilkinson$^1$, A. W. McConnachie$^9$, N. F. Martin$^{10}$, M. B. Davies$^{11}$,
\newauthor
T. J. Bridges$^{12}$\\
$^{1}${Department of Physics and Astronomy, University of Leicester, University Road,
Leicester, LE1 7RH. UK}\\
$^{2}$Research School of Astronomy \& Astrophysics,
The Australian National University,
Mount Stromlo Observatory,
Cotter Road, \\ Weston Creek, ACT 2611, Australia \\
$^{3}${Institute for Astronomy, University of Edinburgh, Royal Observatory,
Blackford Hill, Edinburgh,  EH9 3HJ. UK}\\
$^{4}${%Department of Physics, University of Bristol, Tyndall Avenue, Bristol, BS8 1TL. UK}\\
Astronomisches Rechen-Institut,
Universit{\"a}t Heidelberg,
M{\"o}nchhofstrasse 12-14,
69120 Heidelberg, Germany}\\
$^5$Institute for Astronomy, Swiss Federal Institute of Technology Zurich,
Wolfgang-Pauli-Strasse 27, Building HIT, \\ CH-8093 Zurich, Switzerland\\
$^6$Sydney Institute for Astronomy, School of Physics, A28, The University of Sydney, NSW 2006 Australia\\
$^7$Institute of Astronomy, University of Cambridge, Madingley Road, Cambridge, CB3 0HA. UK\\
$^8$Observatoire Astronomique, Universit\'e de Strasbourg, CNRS, 11 rue de l'Universit\'e,
F-67000, Strasbourg, France\\
$^9$NRC Herzberg Institute of Astrophysics, 5071 West Saanich Road, Victoria, British Columbia, Canada V9E 2E7\\
$^{10}$Max-Planck-Institut f\"ur Astronomie, K\"onigstuhl 17, D-69117 Heidelberg, Germany\\
$^{11}$Lund Observatory, Box 43, SEÐ221 00 Lund, Sweden\\
$^{12}$Department of Physics, Queen's University, Kingston, Ontario K7L 3N6, Canada\\
}

\begin{document}

\date{Accepted . Received ; in original form}

\pagerange{\pageref{firstpage}--\pageref{lastpage}} \pubyear{2009}

\maketitle

\label{firstpage}

\begin{abstract}
We present a structural analysis of halo star clusters in M31 based on
deep {\it Hubble Space Telescope (HST)}
%\hst /ACS
Advanced Camera for Surveys (ACS) imaging.  The clusters in our sample span a range in
galactocentric projected distance from 13 to 100~kpc and thus reside
in rather remote environments. Ten of the clusters are classical
globulars, while four are from the \citet{Huxor05,Huxor08} population
of extended, old clusters.  For most clusters, contamination by M31
halo stars is slight, and so the profiles can be mapped reliably to
large radial distances from their centres.  We find that the
extended clusters are well fit by analytic \citet{King62} profiles
with $\sim20$~parsec core radii and $\sim100$~parsec photometric tidal radii, or by
S\'{e}rsic profiles of index $\sim 1$ (i.e. approximately exponential).
Most of the classical globulars also have large photometric tidal radii in the
range 50--100~parsec, however the King profile is a less good fit in
some cases, particularly at small radii. We find 60\% of the classical
globular clusters exhibit cuspy cores which are reasonably well
described by S\'{e}rsic profiles of index $\sim 2-6$.
Our analysis also
reinforces the finding that luminous classical globulars, with
half-light radii $<10$~parsec, are present out to radii of at least
100~kpc in M31, which is in contrast to the situation in the Milky Way
where such clusters (other than the unusual object NGC 2419) are
absent beyond 40~kpc.
\end{abstract}


\section{Introduction}

In recent years the halo and outer disk of M31 have been 
the subject of several large-area imaging campaigns.
By mapping the distribution of individual resolved stars to large
radii, these have resulted in the discovery of considerable
substructure \citep[e.g.,][]{Ibata01,Ferguson02,Ibata07,McConnachie09},
and the identification of previously unknown 
components in the outer regions of the galaxy
\citep[e.g.,][]{Irwin05,Ibata05,Gilbert06}.  The same surveys have also
been successful in extending the known dwarf galaxy
\citep[e.g.,][]{Zucker04,Martin06,Irwin08,McConnachie08,Martin09,
  Richardson11} and globular cluster systems.

In particular, our group has discovered a number of very remote
globular clusters (GCs) \citep{Huxor04,Martin06,Huxor08, Huxor11}, and
also a population of more extended clusters (ECs), which have
properties similar to the GCs but typical half-light radii nearly an
order of magnitude larger \citep{Huxor05,Huxor08,Huxor11}.  What is
clear from these studies is that the M31 GC system differs from that
of the Milky Way in several respects, in addition to being $\sim3$
times more numerous; notably it includes a significant population of
bright GCs at large galactocentric radius and some of these are very
diffuse \citep{Mackey07, Huxor11}.  By contrast the outer GCs in the
Milky Way, while often moderately extended in structure (typified by
the ``Palomar" clusters),
are predominantly faint\footnote{A notable exception is the unusual,
  remote cluster NGC\,2419, which has been argued may be the core of a
  stripped dwarf spheroidal \citep[e.g.,][]{mackey05,cohen10}.}.  We
note in passing that M31's brighter dwarf spheroidals are also on the
average more extended than the Milky Way's, measured for
example by half-light radius \citep{McConnachie06,Richardson11}, 
although the significance and underlying nature of this discrepancy
remains under debate \citep{collins11,brasseur11}.

Studies of the M31 GC system are important as it possesses the largest
cluster population in the Local Group, and is sufficiently near to us
that clusters can be resolved into stars with \hst.  Several authors
have previously analysed \hst\ images (mainly taken with WFPC2) of
M31's inner GC population
\citep[e.g.,][]{Grillmair96,Holland98,Barmby02,Barmby07,strader11}, finding their
overall structural properties to be similar to those of the Milky-Way
system in the same regime of galactocentric distance.

We have observed a subset of our outer cluster sample with \hst\
Advanced Camera for Surveys (ACS; programme GO-10394, PI Tanvir).  The
targeted clusters included the first four extended clusters
discovered, and ten classical clusters selected because of their large
galactocentric radius.  Photometric analyses of the clusters were
presented in papers by \citet{Mackey06,Mackey07}, which also showed
their spatial distribution around M31, while \cite{Richardson09}
presented an analysis of their surrounding field star populations.  The
clusters are generally old and metal poor, although one classical
globular (GC7) has a relatively high metallicity derived from the
red-giant branch locus of [Fe/H]$\approx-0.7$, and may be
several Gyr younger than the rest of the sample \citep{Mackey07}.

In this paper we consider the structural properties of this sample in
a more detail than hitherto, using 
a combination of surface
photometry and star count analysis.  Such studies provide insight into
the dynamical properties of clusters as well as their origin and
evolution within the parent galaxy's gravitational field.  At large
galactocentric radius we expect only small tidal stresses and also,
because of their long
orbital periods, that visits close to the galactic centre or disk may
have been infrequent in the lifetimes of the clusters.  On the other
hand, it has been argued that many of the outer clusters in the Milky
Way have been accreted from satellite galaxies
\citep[e.g.,][]{Zinn93,Mackey04}, and the situation could be similar in
M31.  Indeed, the same surveys in which these clusters have been found
have also provided evidence of abundant substructure in the M31 halo
and outer disk -- the fossil records of various accretion events --
which we have recently shown show some statistical correlation with
the distribution of outer GCs \citep{Mackey10a}.  Thus the properties
of the outer clusters might alternatively reflect their formation
within dwarf galaxies from which they have been accreted.

Particular goals of this paper are: (a) to compare the structural
properties of the outer GCs with those of the inner galaxy, and the
M31 GC system as a whole to the Milky Way's system; (b) to search for
evidence of tidal disruption as has been seen in a number of MW GCs
\citep[e.g.,][]{Grillmair96}, and might be particularly evident in the
case of the extended clusters. The low density of contaminating field
stars at these large galactocentric radii lends itself to these
studies.

\section{Observations and Data analysis}

The sample of clusters is summarised in Table 1 along with these basic properties and the
name given in the Revised Bologna Catalogue (RBC) of M31 globular 
clusters\footnote{Revised Bologna Catalogue website  {\tt http://www.bo.astro.it/M31/}}.
All observations were made with the \hst /ACS(WFC) using the F606W and
F814W filters.  The clusters were generally located in the middle of
one of the ACS CCDs to minimise the influence of the chip-gap, although in some
instances additional offsets allowed us to avoid bright stars
appearing in the field.  The F814W filter images of all the clusters
are shown in the upper-left panels of Figures 1--14.  In the case of
EC3 and GC6 the two clusters were both available in the same pointing,
and the resulting image (see Figure 3)
illustrates well the striking morphological difference between the two
classes.
Details of the \hst /ACS imaging, the basic reduction steps and
transformation to standard $V$-band (F606W) and $I$-band (F814W)
magnitudes are given in \citet{Mackey06,Mackey07}, along with a
description of the profile-fitting photometric analysis.  We have also
taken the naming scheme, dust extinction values and distances
to 
each cluster from those papers.

\begin{table*}
\centering
 \begin{minipage}{93mm}
\caption{Basic information for our cluster sample.  
We give the name of the cluster used in this work, its entry in the RBC catalogue, its coordinate,
the measured colour excess, $E_{B-V}$, and projected galactocentric radius, $R_{gc}$.}
\begin{tabular}{@{}llllll@{}}\hline\hline
Cluster &  RBC entry & $\alpha_{2000}$ & $\delta_{2000}$ & $E_{B-V}$ & $R_{gc}$ \\ \hline
\vspace{1mm}
EC1   & MCEC1-HEC5  & 00 38 19.5 & +41 47 15 & 0.08 &13.3     \\
\vspace{1mm}
EC2   &  MCEC2-HEC7 & 00 42 55.0 & +43 57 28 & 0.10 &36.8      \\ 
\vspace{1mm}
EC3   &  MCEC3-HEC4 &  00 38 04.6 & +40 44 39 & 0.07 &14.0     \\ 
\vspace{1mm}
EC4   & MCEC4-HEC12 &  00 58 15.4 & +38 03 02 & 0.08 &60.1      \\ 
\vspace{1mm}
GC1   & MCGC1-B520  & 00 26 47.7 & +39 44 46 & 0.09  &46.4     \\ 
\vspace{1mm}
GC2    &  MCGC2-H4 & 00 29 45.0 & +41 13 09 & 0.08 &33.5     \\ 
\vspace{1mm}
GC3    & MCGC3-H5 & 00 30 27.3 & +41 36 20 & 0.11 &31.8      \\ 
\vspace{1mm}
GC4   & B514-MCGC4  & 00 31 09.8 & +37 54 00 & 0.09 &55.2     \\ 
\vspace{1mm}
GC5   & MCGC5-H10  & 00 35 59.7 & +35 41 04 & 0.08 &78.5    \\ 
\vspace{1mm}
GC6   & B298-G021  & 00 38 00.2 & +40 43 56 & 0.09 &14.0       \\ 
\vspace{1mm}
GC7   & MCGC7-H14  & 00 38 49.4 & +42 22 47 & 0.06 &18.2     \\ 
\vspace{1mm}
GC8   & MCGC8-H23  & 00 54 25.0 & +39 42 56 & 0.09 &37.0      \\ 
\vspace{1mm}
GC9   & MCGC9-H24  & 00 55 44.0 & +42 46 16 & 0.15 &38.9   \\ 
\vspace{1mm}
GC10 & MCGC10-H27  & 01 07 26.3 & +35 46 48 & 0.09 &100.0    \\ 
 \hline 
\end{tabular}
\end{minipage}
\end{table*}

\subsection{Surface Photometry}

The first requirement of the analysis is to establish the best
photometric centroid for each cluster.  For the extended clusters and
those classical GCs with low central surface-brightness,
the intensity fluctuations due to individual resolved stars means this
is not straight-forward.  In practice,
we adopted an iterative scheme, beginning with a centroid defined by
the simple first moment of the light distribution and hence finding
the best-fitting King function from a first pass (see below) analysis
of the 1D profile.  This analytic function was itself then fitted to
the 2D spatial distribution of stars to obtain an improved estimate of
the centroid, and hence an improved profile fit and so-on until
convergence.
This approach works well for all the clusters, with the possible
exception of the sparsest, EC4, for which the maximum likelihood fit
finds a number of plausible local maxima (within $\sim2$\,arcsec) which are not much less
likely than the best solution.

Measurements of the classical clusters with the {\sc
  Iraf}/ellipse\footnote{IRAF is distributed by the National Optical
  Astronomy Observatory, which is operated by the Association of
  Universities for Research in Astronomy (AURA) under cooperative
  agreement with the National Science Foundation.}  task finds no
significant ellipticity in the inner regions, whilst the extended
clusters also appear to be roughly circular, within the limits of the
noise introduced in the light distribution by individual resolved
stars.
For each cluster we therefore summed the integrated light in circular
annuli around the cluster centre, and subtracted an estimate of the
background using areas of the images well beyond the apparent cluster
extent (typically a distance of about 1 arcmin defined the inner
radius of the background region).  Only particularly bright stars and
galaxies were masked from the background estimate (the same sources
are also masked from the cluster apertures).
It is worth noting at this point that if any of the clusters extend at
low levels over a majority of the ACS fields, then it would lead us to
overestimate the background level, and underestimate, for example, the
photometric tidal radius.

The spacing of the boundaries between the annuli is logarithmic,
although the number of annuli used for each cluster was chosen,
dependent on the size and brightness of the cluster, to ensure good
signal-to-noise in most bins (in particular, relatively few but large
bins were considered most appropriate for the extended clusters).

\subsection{Star Counts}

For the classical globular clusters at larger radii, and the extended
clusters at all radii, these integrated light surface photometry
measurements become noisy due to contamination: the result of light
from faint background galaxies and stars in M31 and the MW halo (as
noted above, the brighter stars and galaxies in the images were masked
out to reduce their effect on the star counts).  To extend the
measured profiles we performed star counts using the following
procedure: \\
\noindent {\bf Step 1:} a colour-magnitude diagram (CMD, see
lower-left panels of Figures 1--14) was created for each cluster.
Bold symbols represent stars from an annulus judged by eye to be
clearly within the main body of the cluster but not too severely
crowded (typically out to about 10 to 20 arcsec, depending on the
cluster), and therefore are likely to be dominated by cluster stars.
Smaller symbols represent stars from the same region of the image that
was used to calculate the sky background,
which are very likely to be almost entirely sources unassociated with
the cluster.  Stars in the outer parts of the cluster were excluded
from this plot since they could not be considered either good
candidates for membership or likely contaminants, simply on the basis
of spatial position.
For this reason the primary references for the cluster CMDs remain \citet{Mackey06,Mackey07},
in which the typical photometric errors are also presented.
\\
\noindent {\bf Step 2:} 
regions on the CMD were then chosen separately for each cluster (shown
as outlines in the figures) in order roughly to maximise the number of
cluster stars, whilst minimising background contamination.  The more
remote clusters with very low contamination therefore benefitted from
much looser constraints in this regard.  Nevertheless, it is important
to note that effective background contamination varies considerably
from cluster to cluster, not only because the clusters nearer to M31
have a higher density of M31 halo and disk stars in the field, but
also because in some cases, driven by metallicity differences, the
loci of the cluster red giant branch (RGB) and horizontal branch (HB)
were less distinct from those of the M31 halo than in other cases.
Clusters EC3 and GC6, which are both located in the same ACS field,
suffered most from a high density of contaminating field stars.  \\
\noindent {\bf Step 3:} stars from within these CMD regions were the
only ones, henceforth, considered in the star count analysis.  Their
numbers were summed in the same annuli as were used for the surface
photometry. Residual contamination within these annuli, by
point-sources unrelated to the cluster, was removed
statistically by subtracting an appropriately scaled background count
determined, again, over the area of the chip far from the cluster.

The spatial distribution of stars within the selected regions of each
CMD is shown in the upper-right panel of Figures 1--14.  The different
regions of the CMD used to measure the star counts are colour-coded to
highlight different stellar populations (usually, horizontal branch,
and upper and lower red giant branch).  Given that the range of
stellar mass from the RGB to the HB is not large, we do not expect any
significant difference in these distributions, except that which
results from incompleteness due to crowding setting in at larger radii
for the fainter stars.  In these panels we also plot for reference the
direction of M31 and the King-profile core and photometric tidal radii (see below)
of each cluster.  Note the linear scale in parsec is appropriate to
the measured distance of each cluster \citep{Mackey06,Mackey07}.

\subsection{Cluster Profiles}

To construct the final, circularly-averaged intensity profiles, the star count density was
scaled to the $V$-band surface photometry in an annular region, the
inner and outer radii of which were selected by eye for each cluster
to define a zone in which the surface photometry should still be
reliable, but the star counts not significantly affected by
incompleteness.
For the classical globulars, this was typically a region from a few
arcsec to about 15 arcsec from the cluster centre.  In effect, this
means that core radii are determined largely from the integrated light
measurements and photometric tidal radii largely from the star counts.  In the
case of extended clusters the profiles reported are those solely based
on the star counts, although again these are normalised to the
measured surface photometry in an aperture roughly corresponding to
the core region.

The resulting radial profiles (extinction-corrected) are shown in
lower-right panels of Figures 1--14.  The error bars for the star
counts were determined from Poisson statistics (including the error
introduced by the background subtraction step), whereas those for the
surface photometry were obtained by dividing each annulus into eight
segments and determining the variance of surface brightness between
these.  This latter approach should account for the extra noise, over
and above photon shot noise plus detector noise etc., which is due to
the graininess introduced by resolved stars; although for very small
radii the segments are typically so small that the variance is likely
to be underestimated.  In the overlap region, in which the plotted
points are a weighted average of the surface photometry and (scaled)
star counts, the error is taken as an average of the error on each
(this was chosen since the uncertainties should be well correlated).

Due to the magnitude limit of the observations, the star counts only
sample the most luminous stars in the clusters.  The integrated light
analysis is also dominated by the light from the brightest, most
massive stars, and particularly in the intermediate radii where we
normalise the star counts to the surface photometry, it is reasonable
to assume dynamical mass segregation does not affect the results.  At
small radii, particularly in any post core-collapse clusters, sinking
of more massive stars (including stellar remnants) to the cores will
modify both the luminosity function and mass-to-light ratio.
The radial dependence of $V-I$ colour, determined from surface
photometry in annuli around each cluster (see bottom plots in lower
right panels of Figures 1--14), confirms there is no evidence of any
significant trend on the scales we probe.

\begin{figure*}
\begin{center}
\includegraphics[width=16.8cm,angle=0]{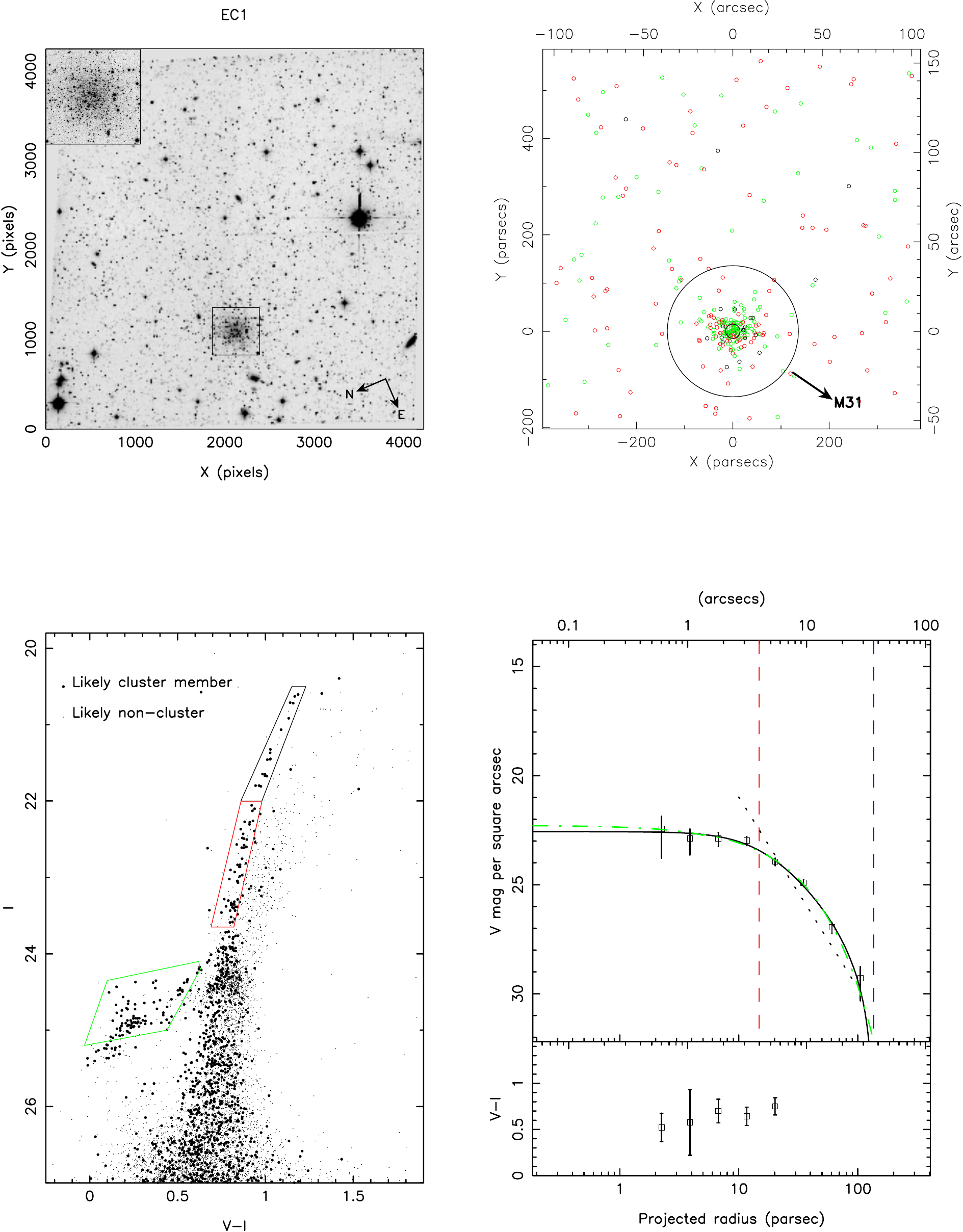}
\caption{ Results for extended cluster EC1. Upper left: ACS/F814W
image of the field (inset shows zoom-in on cluster); Lower left: colour magnitude diagram of
point-sources in the field, with likely cluster stars, which is to say
those clearly within the main body of the cluster, in larger symbols.
The points labelled as likely non-cluster are from areas of the chip
well away from the cluster.  Outlined regions are areas chosen to be
dominated by cluster populations; Upper right: the spatial
distribution of stars from the (colour-coded) regions outlined in the CMD.  The two
circles indicate the core radius and tidal radius determined from the
fitted King profile.  The arrow indicates the direction of the centre
of M31. Lower right: radial profile determined from star counts scaled
to match the extinction-corrected $V$-band surface photometry in the
inner regions, along with the best fitting King profile (solid curve), and S\'ersic
profile (dashed curve).  Dashed vertical lines are the core and tidal
radii from the King fit. The V-I surface photometry colour is shown
below.}
\end{center}
\end{figure*}

\begin{figure*}
\begin{center}
\includegraphics[width=16.8cm,angle=0]{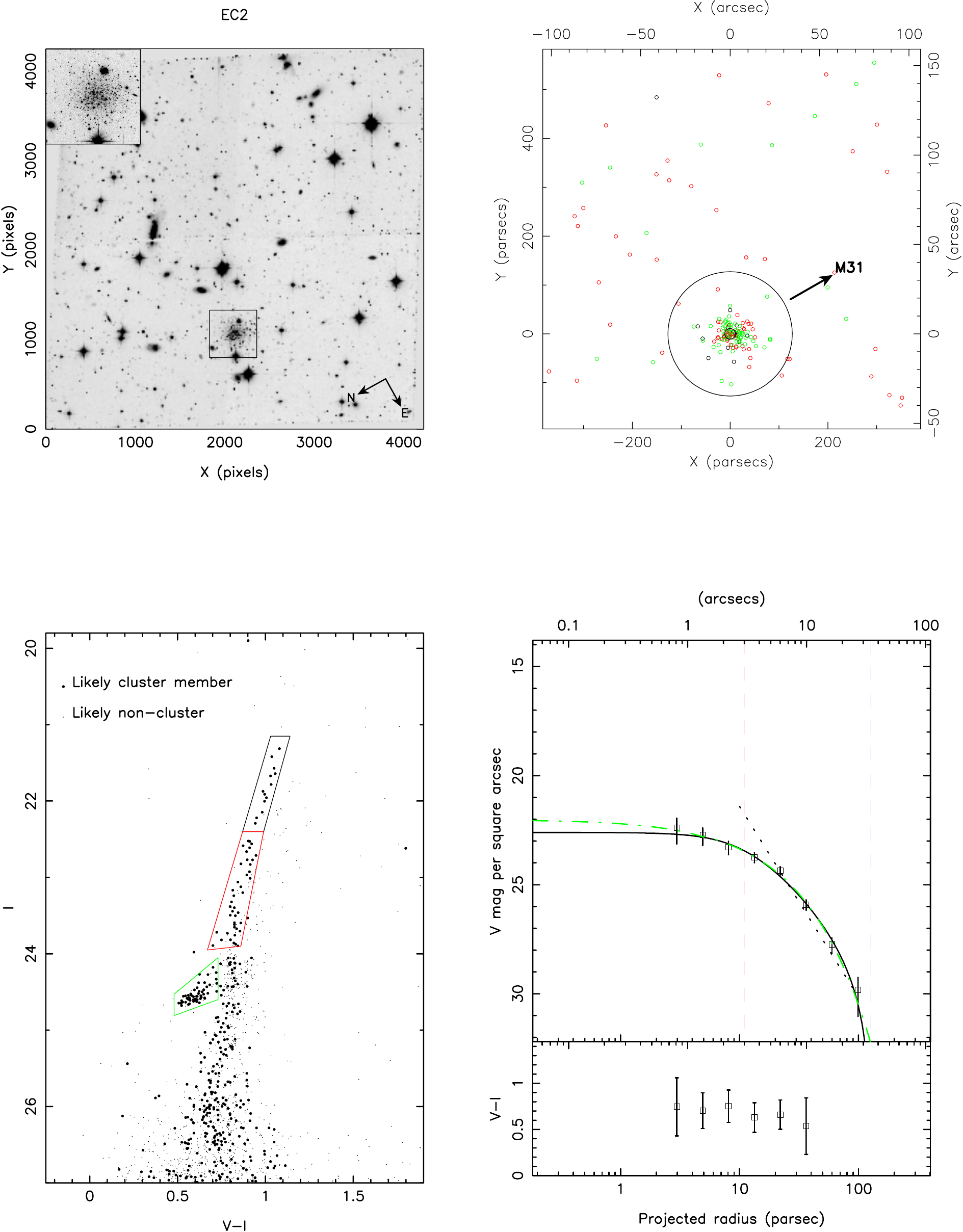}
\caption{ Results for extended cluster EC2. Panels are as in Figure
1.  There is some indication to the eye of a deviation from circular
symmetry in the outer parts of this cluster (specifically the distribution seems to be 
extended in the positive $x$-direction).  As discussed in the text, our statistical
analysis  suggests this is not significant.}
\end{center}
\end{figure*}

\begin{figure*}
\begin{center}
\includegraphics[width=16.8cm,angle=0]{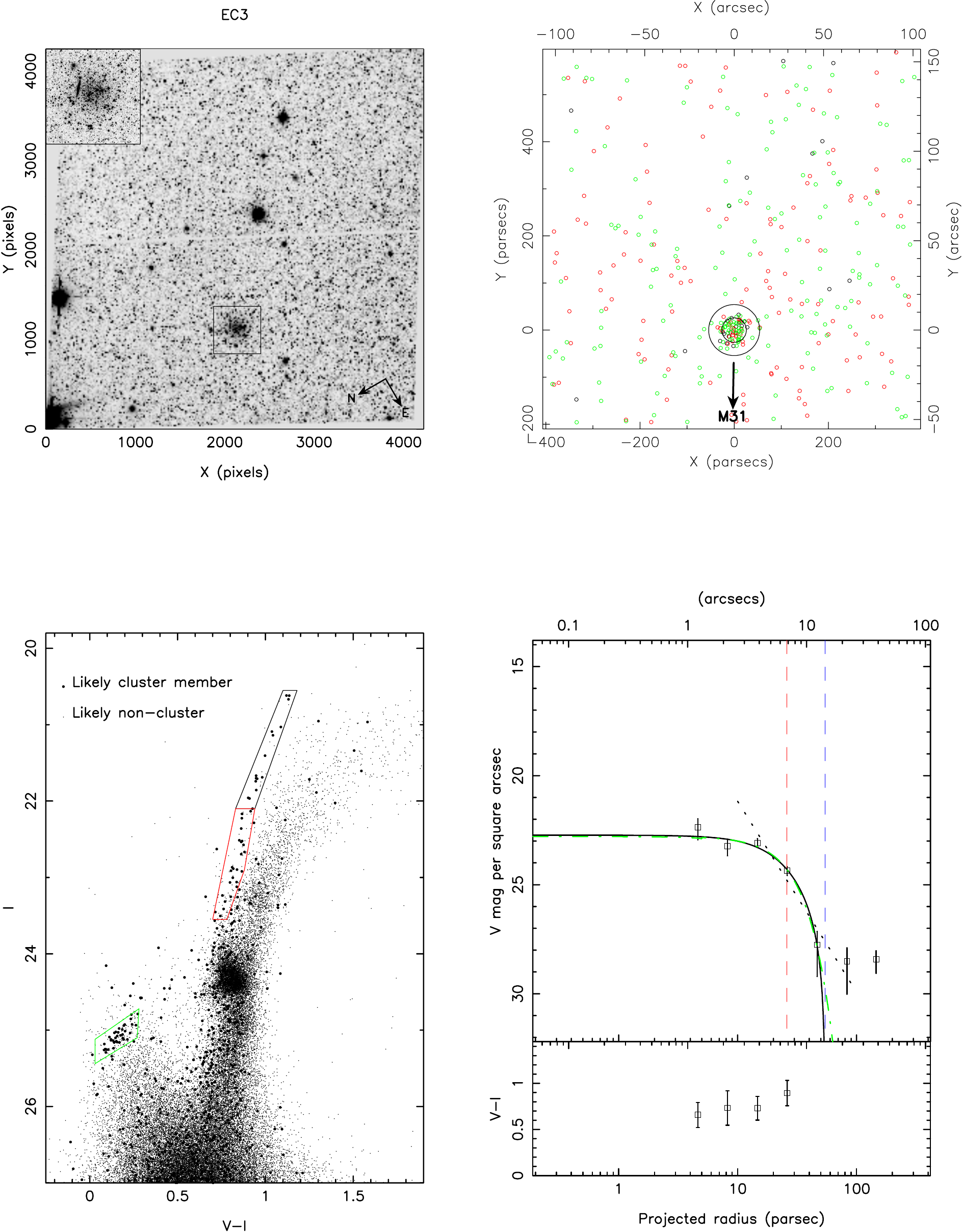}
\caption{ Results for extended cluster EC3. Panels are as in Figure 1.
Note that GC6 appears in the same frame, but is masked out from the
background region.}
\end{center}
\end{figure*}

\begin{figure*}
\begin{center}
\includegraphics[width=16.8cm,angle=0]{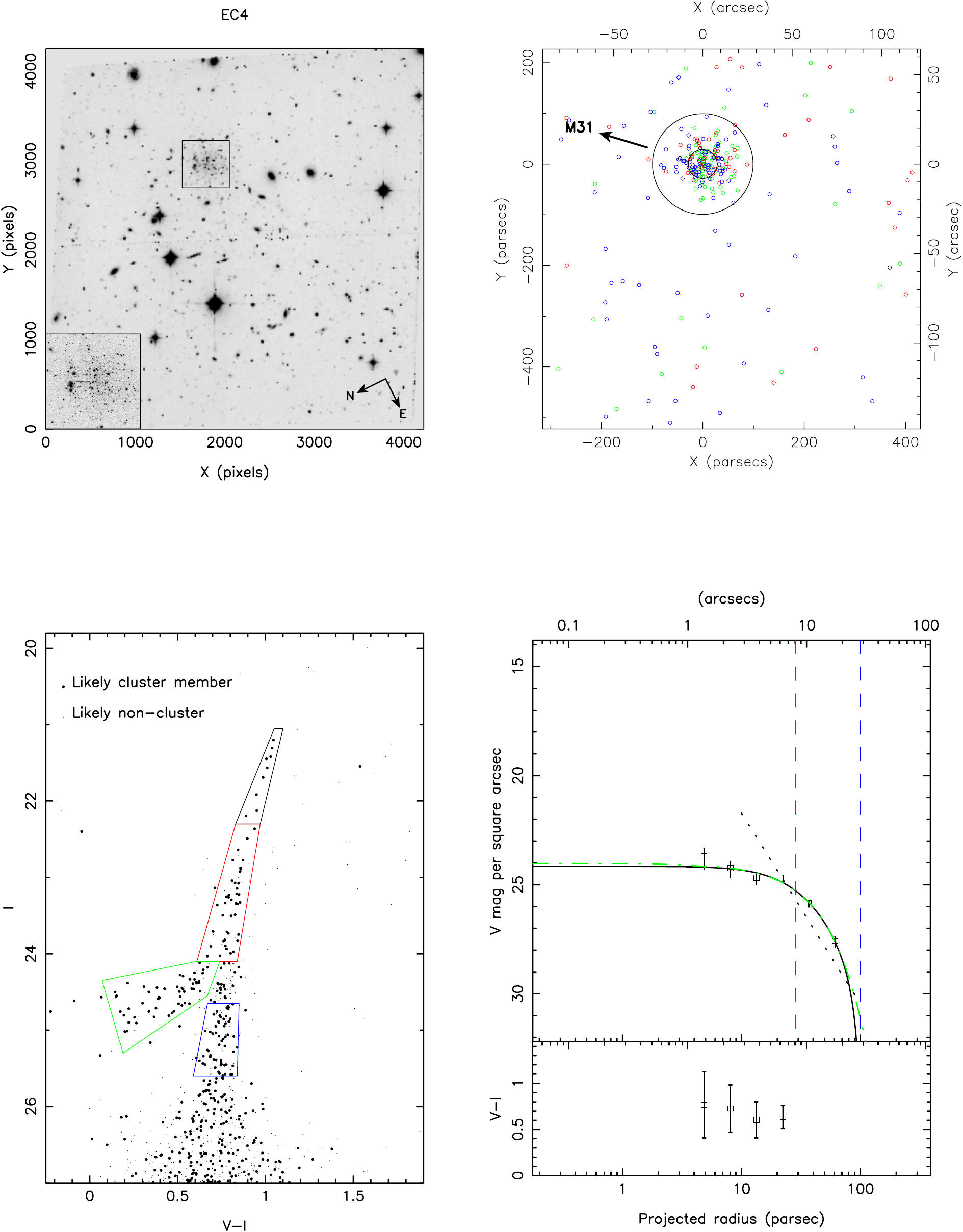}
\caption{ Results for extended cluster EC4. Panels are as in Figure
1. Curiously, to the eye, there appears to be an asymmetry in the distribution
of horizontal branch stars in this cluster, with an excess in the positive $x$-direction.  As discussed in the text, this
appears to be moderately significant statistically, but would be hard to
explain by dynamical processes, so we conclude is most likely an
unusual chance occurance.}
\end{center}
\end{figure*}

\begin{figure*}
\begin{center}
\includegraphics[width=16.8cm,angle=0]{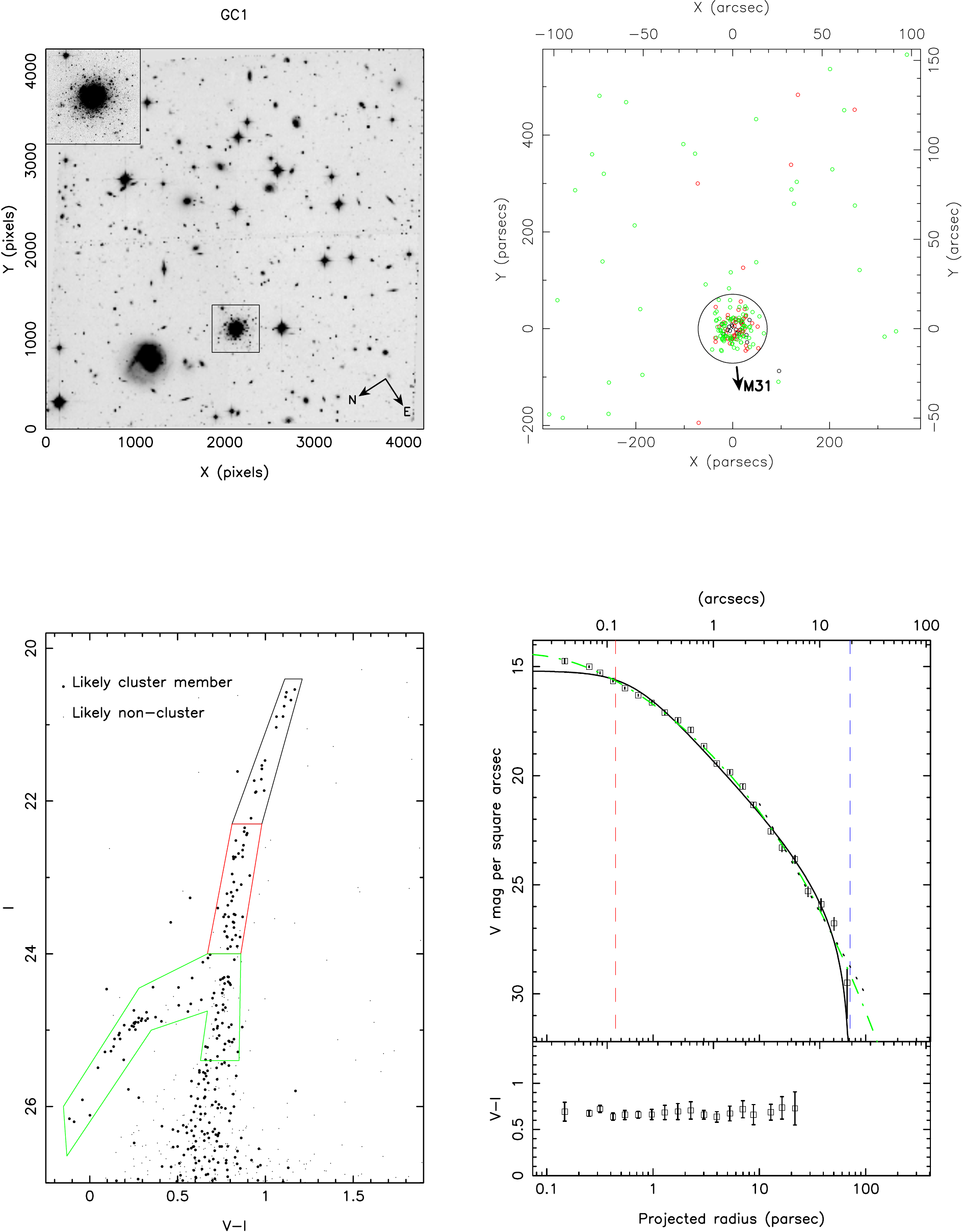}
\caption{ Results for classical globular cluster GC1. Panels are as in
Figure 1 with the exception of the (bottom right panel) radial profile
which is based on a combination of surface photometry in the inner
regions and (scaled) star counts in the outer regions.  The overlap
range is fixed by eye for each cluster.  In addition to the
King-profile fit (solid curve), we also plot the best fitting S\'ersic
profile (dashed curve).}
\end{center}
\end{figure*}

\begin{figure*}
\begin{center}
\includegraphics[width=16.8cm,angle=0]{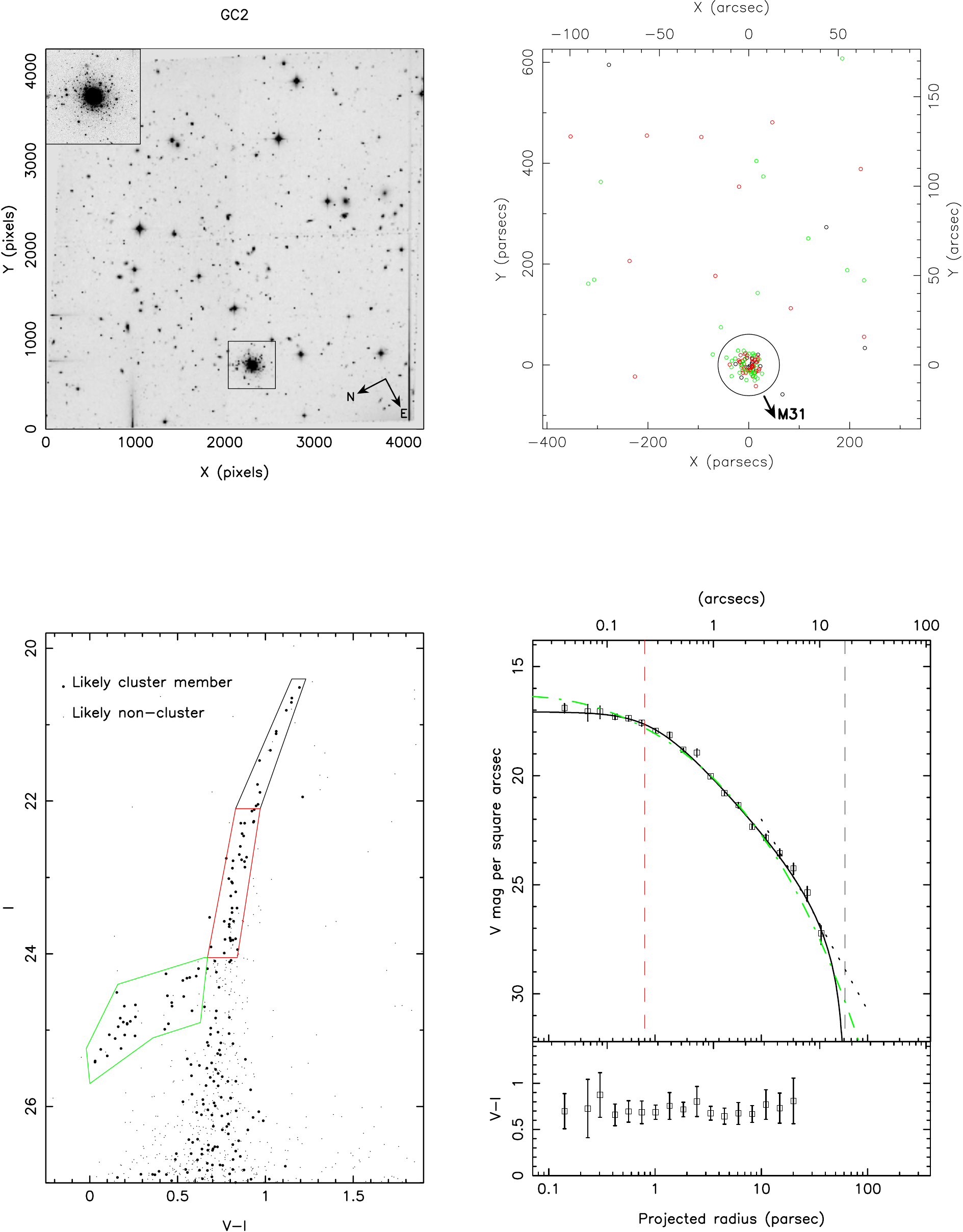}
\caption{ Results for classical globular cluster GC2. Panels are as in
Figure 5.}
\end{center}
\end{figure*}

\begin{figure*}
\begin{center}
\includegraphics[width=16.8cm,angle=0]{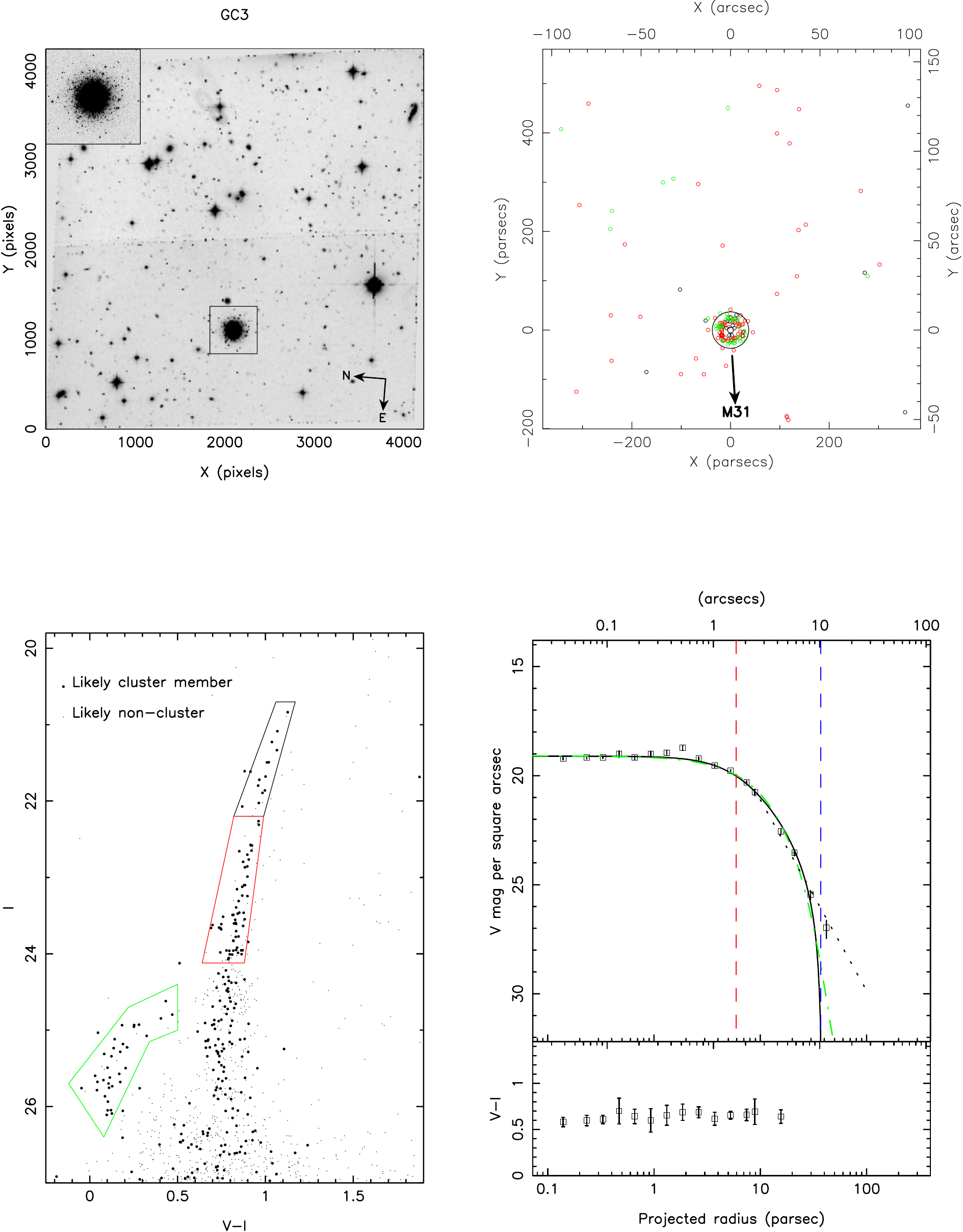}
\caption{ Results for classical globular cluster GC3. Panels are as in
Figure 5.  The annular images towards the top of the image are ghost
images of a bright star.}
\end{center}
\end{figure*}

\begin{figure*}
\begin{center}
\includegraphics[width=16.8cm,angle=0]{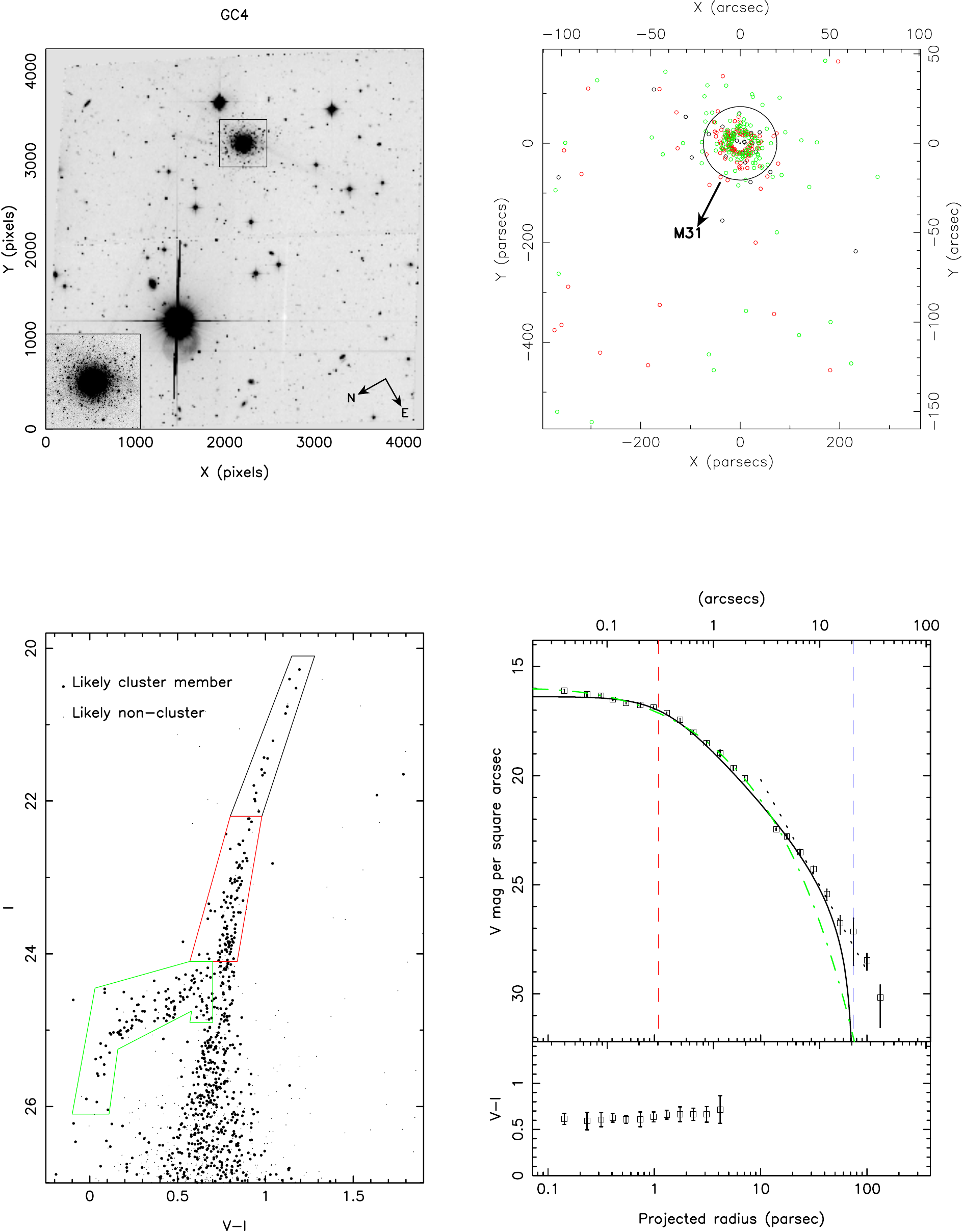}
\caption{ Results for classical globular cluster GC4. Panels are as in
Figure 5.  Note that it was not possible to reliably determine the
colour of the cluster at larger radii due to some ghosting and flaring in the
images (presumably from the bright star) which affected the surface
photometry around the cluster.  This in itself is a good illustration
of the advantages of using star counts to map the outer regions of
clusters, rather than integrated photometry.  
This is the only cluster with significant numbers of stars beyond the fitted
tidal radius, and may indicate an overflow of stars, as discussed in the text.
}
\end{center}
\end{figure*}

\begin{figure*}
\begin{center}
\includegraphics[width=16.8cm,angle=0]{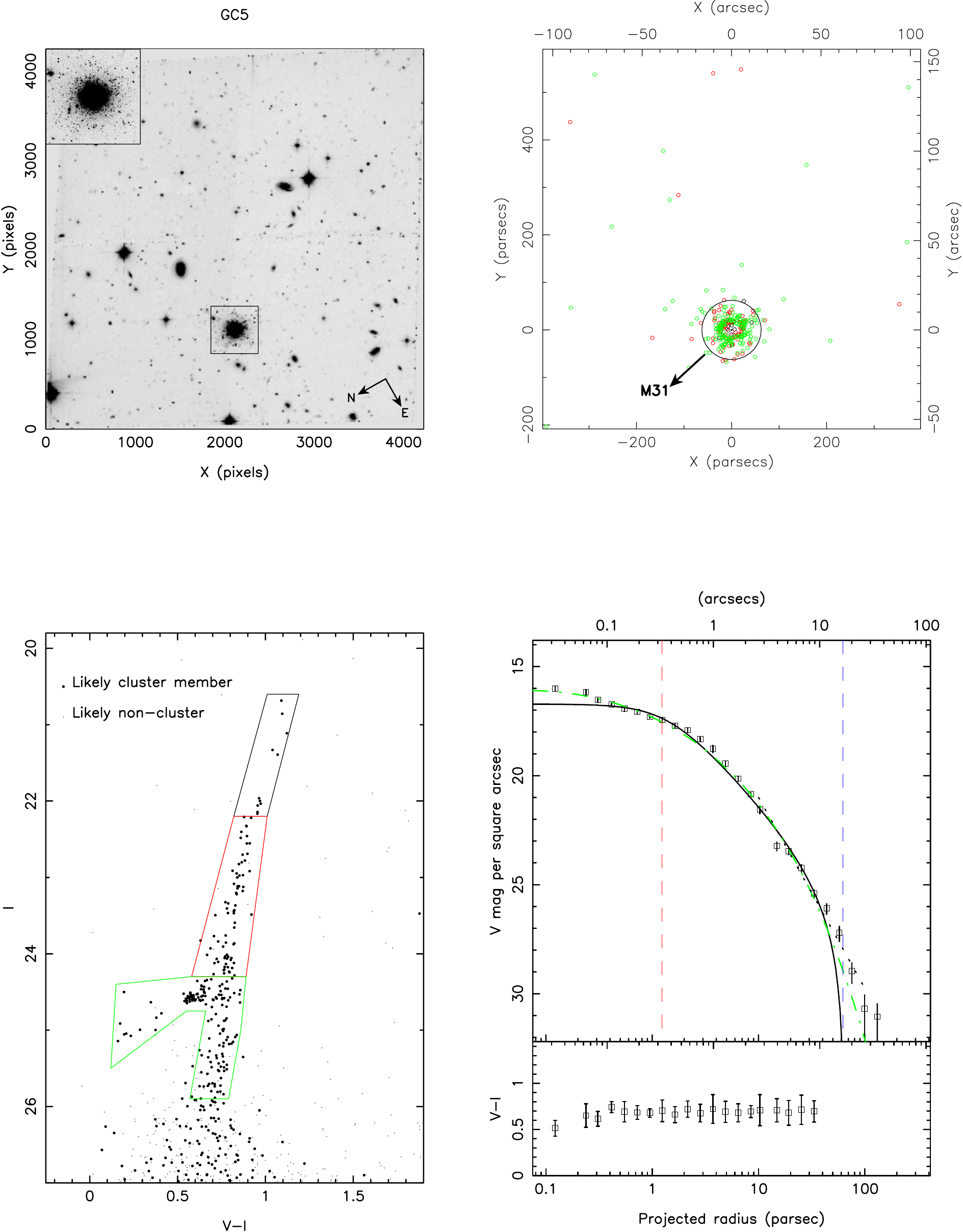}
\caption{ Results for classical globular cluster GC5. Panels are as in
Figure 5.}
\end{center}
\end{figure*}

\begin{figure*}
\begin{center}
\includegraphics[width=16.8cm,angle=0,clip]{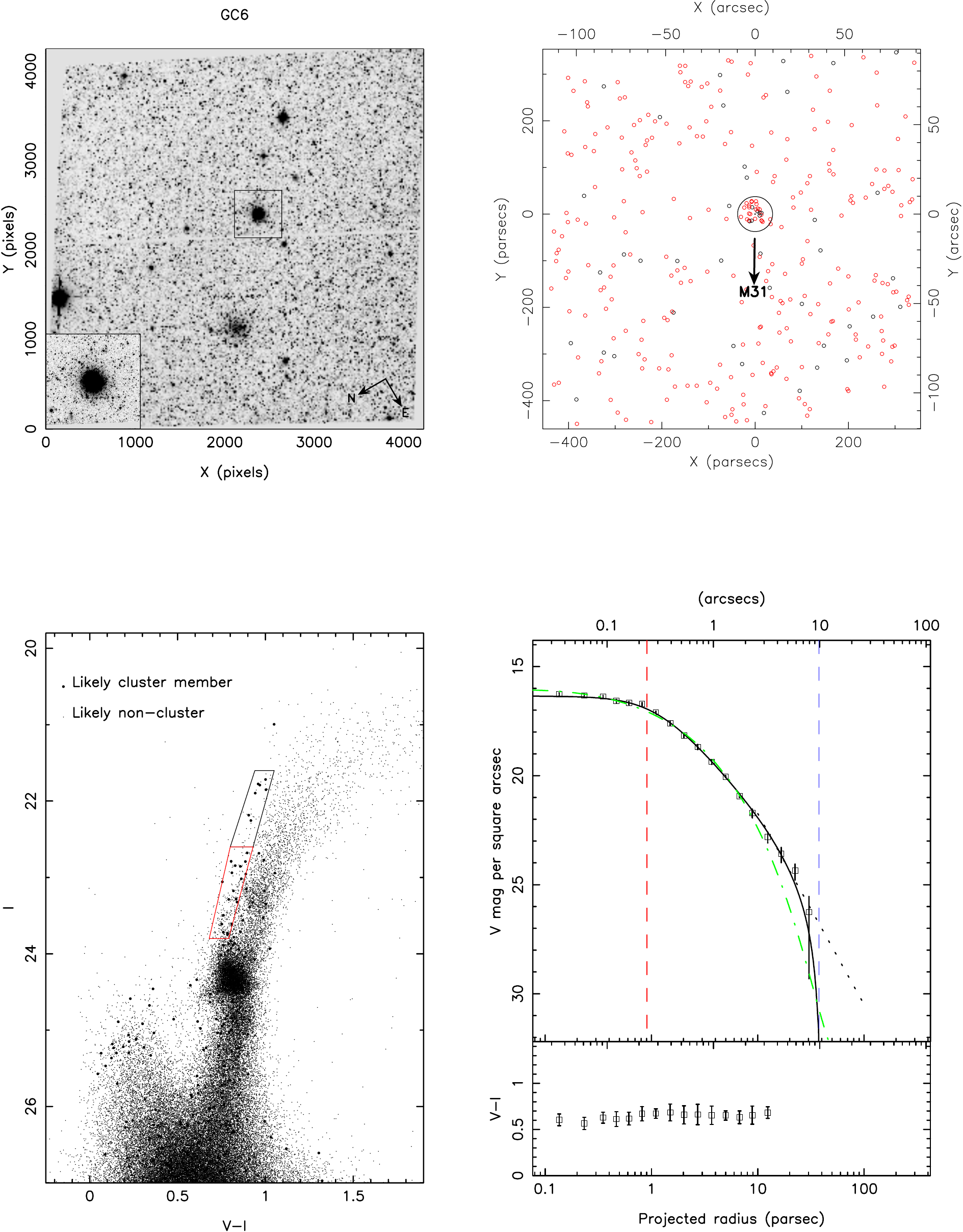}
\caption{ Results for classical globular cluster GC6. Panels are as in
Figure 5.  Note that EC3 appears in the same frame, but is masked out
from the background region.}
\end{center}
\end{figure*}

\begin{figure*}
\begin{center}
\includegraphics[width=16.8cm,angle=0]{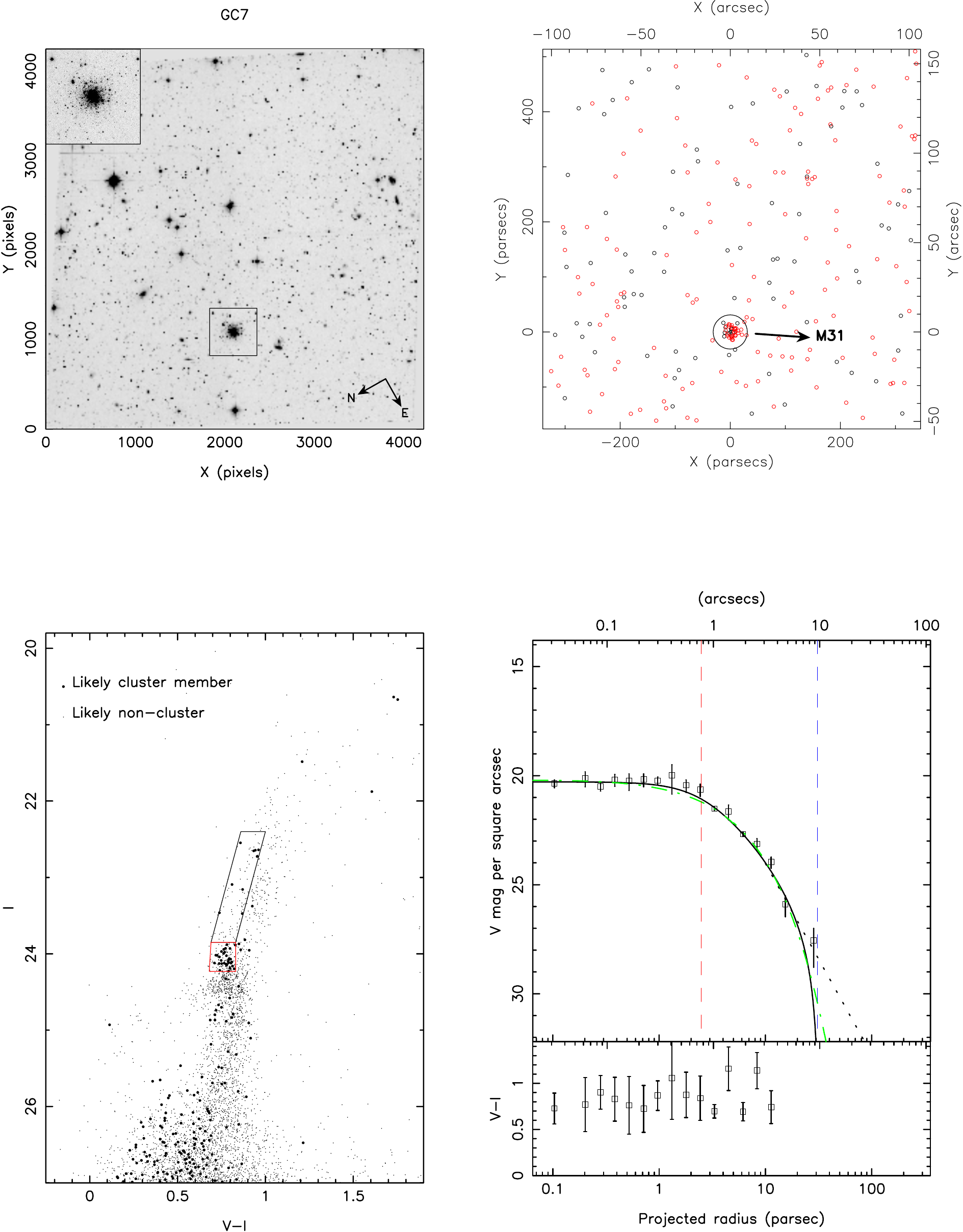}
\caption{ Results for classical globular cluster GC7. Panels are as in
Figure 5.}
\end{center}
\end{figure*}

\begin{figure*}
\begin{center}
\includegraphics[width=16.8cm,angle=0]{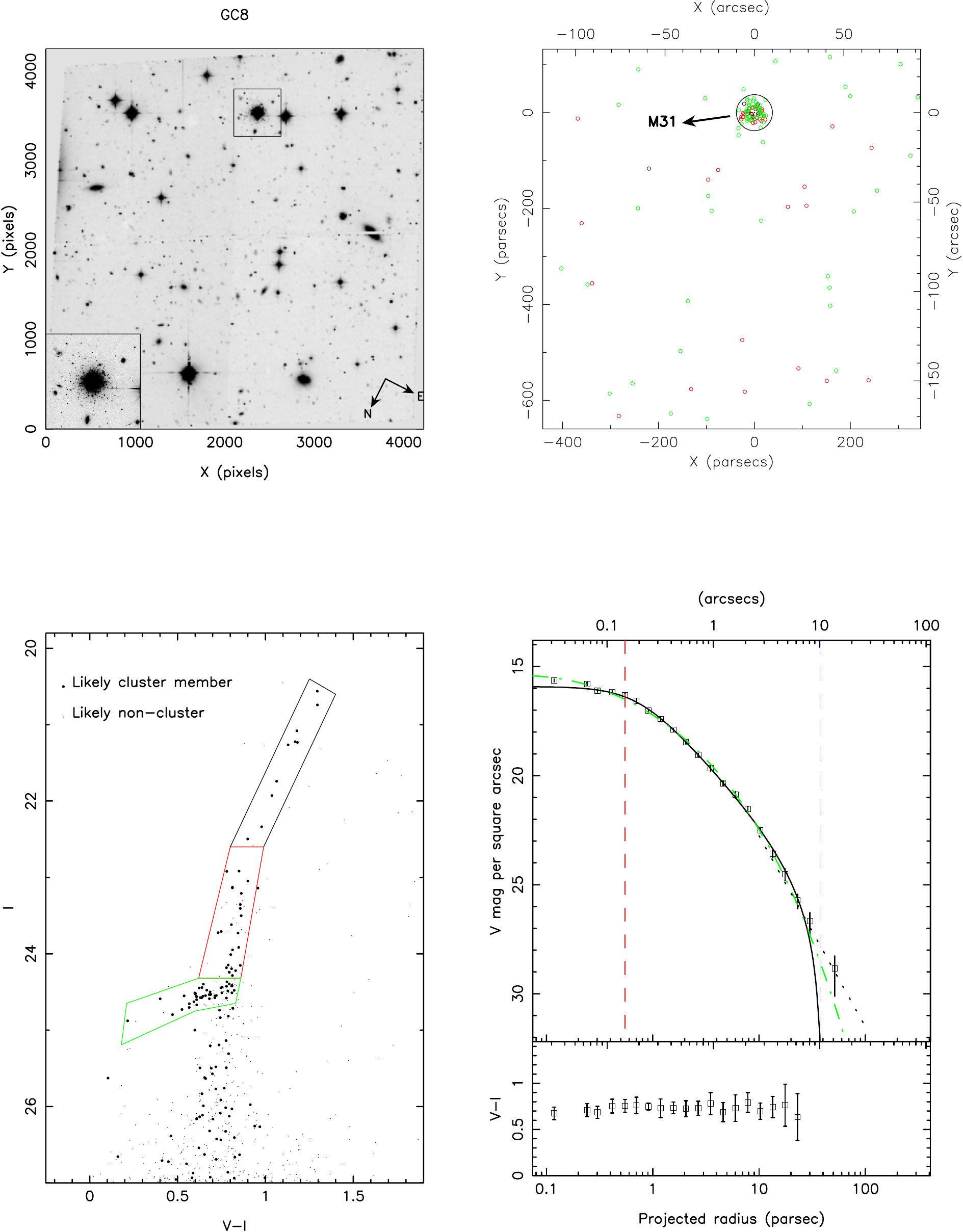}
\caption{ Results for classical globular cluster GC8. Panels are as in
Figure 5.}
\end{center}
\end{figure*}

\begin{figure*}
\begin{center}
\includegraphics[width=16.8cm,angle=0]{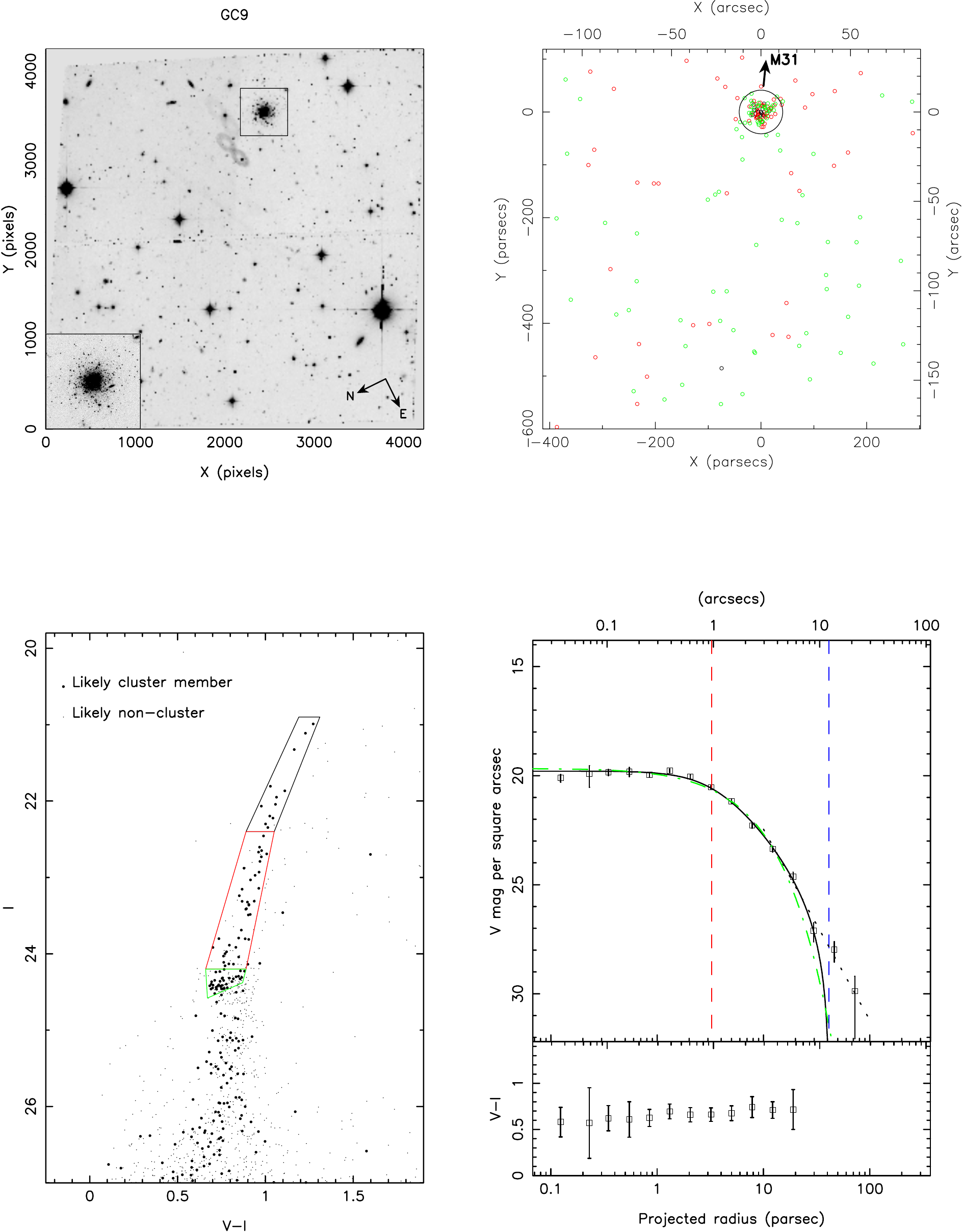}
\caption{ Results for classical globular cluster GC9. Panels are as in
Figure 5. The annular images towards the top of the image are ghost
images of a bright star.}
\end{center}
\end{figure*}

\begin{figure*}
\begin{center}
\includegraphics[width=16.8cm,angle=0]{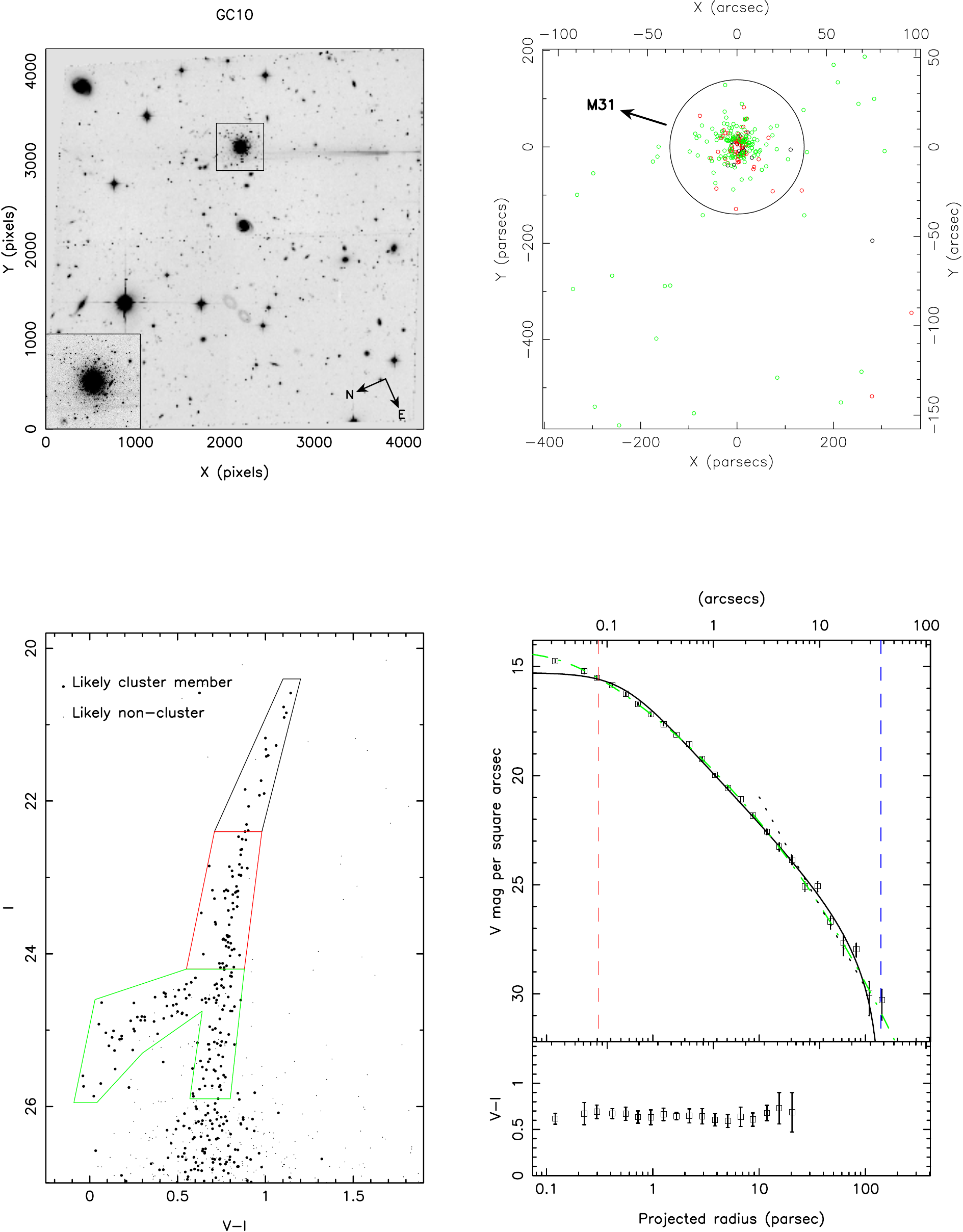}
\caption{ Results for classical globular cluster GC10. Panels are as
in Figure 5. Both flares and ghost images due to bright stars
off the frame are visible, but do not affect the analysis.}
\end{center}
\end{figure*}

\section{Results}

We fitted the radial  profile of each cluster
with an empirical
\citet{King62} model of the form:

\begin{equation}
I(R)=I_0\left[{1\over{\sqrt{1+\left({R\over R_c}\right)^2}}}-{1\over{\sqrt{1+\left({R_t\over R_c}\right)^2}}}\right]^2 \quad ; R\le R_t
\end{equation}

\noindent
where $I(R)$ is the surface brightness at projected radius $R$, $I_0$
is a normalising factor (which is approximately the central surface
brightness when $R_t>>R_c$), $R_c$ the core radius (again,
approximately the half-width at half-maximum in the same limit), and
$R_t$ is the photometric tidal cut-off radius.  It is worth emphasising that
$R_t$ does not necessarily correspond closely to the current physical (dynamical) tidal radius, even when a
King model provides a good fit to the light profile, since, for example, it could reflect past
tidal curtailment when the cluster's orbit passed closer to the galaxy, or be influenced
by stars in the process of escaping the cluster.

The best fit was
determined using a Markov-Chain Monte-Carlo algorithm to find the
maximum likelihood model parameters, taking the estimated photometric
flux errors to be Gaussian distributed.  These fits are the solid
curves shown in the lower-right panels of Figures 1 -- 14.  In
addition, the data were similarly fitted with \citet{Sersic68}
profiles:

\begin{equation}
I(R)=I_0~{\rm exp}\left[ -\left({R\over R_S}\right)^{1/n}\right]
\end{equation}

\noindent
where $R_S$ is a scale radius and $n$ is the shape index.
The S\'ersic profile (dashed curves shown in the same panels) has no
tidal cut-off, but better describes steeper, more cuspy central light
distributions seen in some clusters.

Note that the models were convolved with \hst /ACS point spread
function (PSF) -- the functional form of the PSF being taken from
\citet[ see their Equation 4 and Table 3]{Barmby07}.  Although the
difference is only small, it provides a more correct comparison with
the data in the cluster cores.  Of course, the pixel binning
contributes further to the smoothing of light in the very central
regions, and imposes another limit to the information we can expect to
find, particularly for clusters with central cusps.

For the extended clusters, the low density of stars precludes reliable
measures of star counts or surface brightness in the central couple of
arcseconds since the annular bins there are such that a single
resolved star (or the absence of one) makes a significant difference
to the measured flux.  For the same reason, estimating flux
uncertainties for small regions in the extended clusters is less
reliable.  Thus, although both the King and S\'ersic profile fits are
good, we cannot make firm statements about how flat the central light
distribution is in each such object.

We also calculated model-independent (hereafter referred to as
``empirical'') estimates of half-light radius and absolute magnitude.
For the extended clusters this was calculated directly by summing the
star counts and scaling these to the surface photometry in the central
regions.  For the classical globulars we combined surface photometry
in the inner regions and (scaled) star counts in the outer regions, to
give an effective brightness distribution, which was again summed to
determine these parameters.

Numerical results of our analysis, including $1\sigma$
confidence
intervals for the fitted parameters,
are summarised in Table  2. The galactocentric radii are projected on the sky from the galaxy centre, and we take the
distance to M31 to be 785\,kpc \citep{mcconnachie05}.  The empirical
absolute magnitudes and half light radii presented here can be
considered to be more precise than the ground-based estimates in
\citet{Huxor05}, and the preliminary estimates using the same data
tabulated in \citet{Mackey06,Mackey07}.  The agreement is generally
good, with a slight systematic tendency to brighter magnitudes and
smaller radii here.  We note that EC3 is a particularly difficult case
due to the
high background of contaminating field stars, and formally the core radius is found to be
only a factor of $\sim2$ smaller than the tidal radius, although there may be a
slight excess of stars beyond the tidal radius.

GC4 was observed independently by \citet{Federici07}, also with
\hst/ACS.  Their derived parameters from a King model fit, $M_V=-9.1$,
$R_c=1.4$\,pc and $R_h=5.4$\,pc, are in good agreement with those found
in our analysis.

\begin{table*}
\centering
 \begin{minipage}{170mm}
\caption{Results for the extended and  classical globular clusters. 
 For each we give the empirical magnitudes and half-light radii; half-light, core and tidal radii, and central
surface brightness for King profile fits; half-light, core and tidal radii, and central
surface brightness for S\'ersic profile fits; estimated pericentric distance from M31 (see text for more details)}
\begin{tabular}{@{}llllllllllllll@{}}\hline\hline
Cluster &    \multicolumn{2}{c}{Empirical} &&   \multicolumn{4}{c}{King}  &&    \multicolumn{4}{c}{S\'ersic}     &  $R_{\rm peri}$    \\ \cline{2-3}\cline{5-8}\cline{10-13}
              &      $M_V$            & $R_h$ && $R_c$ & $R_h$ & $R_t$ & $\mu_{V0}$ &&  log$_{10}(R_S/$pc) & $R_h$  & n       & $\mu_{V0}$& (kpc)\\ 
              &                               &   (pc)    &&    (pc)   &   (pc)    &    (pc)   &                  &&    & (pc)  &     & & \\ \hline
\vspace{1mm}
EC1       &   -7.68 &   33.2  &&   $14.8^{ +3.5}_{-3.3}$ & 24.2 & $ 136^{+22}_{-34}$ & $22.55^{+.31}_{-.18}$  &&    $+1.12^{+.15}_{-.16}$ & 24.4 & $  1.1^{+0.1}_{-0.2}$ & $22.24^{+.49}_{-.29}$ & 20.3\\
\vspace{1mm}
EC2        &   -7.03 &   24.9  &&   $10.9^{+ 2.9}_{- 3.0}$ & 20.1 & $ 127^{+ 28}_{- 36}$ & $22.59^{+.39}_{-.26}$  &&    $+0.86^{+.27}_{-.23}$ & 19.9 & $  1.3^{+0.1}_{-0.4}$ & $21.94^{+.92}_{-.24}$ & 25.7  \\ 
\vspace{1mm}
EC3        &   -7.45 &   24.6  &&   $26.0^{+13.3}_{- 4.9}$ & 18.3 & $  54^{+  5}_{-  8}$ & $22.72^{+.29}_{-.19}$  &&    $+1.34^{+.09}_{-.04}$ & 18.1 & $  0.5^{+0.1}_{-0.2}$ & $22.78^{+.42}_{-.20}$  & 7.2 \\ 
\vspace{1mm}
EC4        &   -6.68 &   33.2  &&   $28.5^{+ 4.5}_{-11.0}$ & 27.7 & $  99^{+ 27}_{- 35}$ & $24.14^{+.22}_{-.25}$   &&   $+1.41^{+.10}_{-.13}$ & 28.1 & $  0.7^{+0.2}_{-0.2}$ & $24.01^{+.38}_{-.33}$ & 25.8  \\ 
\vspace{1mm}
GC1      &    -8.91 &    3.5  &&    $ 0.44^{+ .02}_{- .01}$ &  2.8 & $  71^{+ 5}_{- 5}$ & $14.68^{+.05}_{-.04}$  &&  $-2.73^{+.10}_{-.22}$ &  3.0 & $  3.7^{+0.1}_{-0.1}$ & $10.93^{+.20}_{-.21}$ & 7.1 \\ 
\vspace{1mm}
GC2      &   -7.83 &    4.0  &&   $ 0.80^{+ .04}_{- .04}$ &  3.6 & $  61^{+ 7}_{- 6}$ & $16.81^{+.09}_{-.09}$  &&   $-1.44^{+.23}_{-.32}$ &  3.5 & $  2.8^{+0.1}_{-0.3}$ & $14.46^{+.45}_{-.18}$ & 8.2  \\ 
\vspace{1mm}
GC3      &  -8.68 &    9.1  &&    $ 5.92^{+ .22}_{- .17}$ &  7.9 & $  37^{+ 1}_{- 1}$ & $19.08^{+.02}_{-.02}$  &&   $ +0.83^{+.01}_{-.01}$ &  8.0 & $  0.8^{+0.0}_{-0.0}$ & $19.06^{+.03}_{-.03}$ & 3.0 \\ 
\vspace{1mm}
GC4      &   -8.74 &    3.5  &&    $ 1.09^{+ .03}_{- .04}$ &  4.6 & $  74^{+ 7}_{- 7}$ & $16.19^{+.03}_{-.03}$  &&   $-0.42^{+.07}_{-.07}$ &  4.3 & $  1.9^{+0.1}_{-0.1}$ & $15.21^{+.11}_{-.08}$ & 8.0 \\ 
\vspace{1mm}
GC5      &   -8.90 &    5.1  &&   $ 1.23^{+ .06}_{- .05}$ &  4.5 & $  62^{+  5}_{-  5}$ & $16.55^{+.05}_{-.05}$  &&   $-0.81^{+.11}_{-.11}$ &  4.9 & $  2.3^{+0.1}_{-0.1}$ & $14.89^{+.19}_{-.12}$ & 6.0 \\ 
\vspace{1mm}
GC6      &    -8.50 &    3.6  &&    $ 0.91^{+ .03}_{- .04}$ &  3.0 & $  38^{+  5}_{- 4}$ & $16.10^{+.04}_{-.04}$  &&  $-0.33^{+.05}_{-.07}$ &  2.7 & $  1.6^{+0.1}_{-0.1}$ & $15.37^{+.09}_{-.09}$ & 3.2  \\ 
\vspace{1mm}
GC7      &    -6.25 &    5.4  &&    $ 2.50^{+ .18}_{- .22}$ &  4.7 & $  31^{+ 5}_{- 5}$ & $20.22^{+.10}_{-.09}$  &&    $ +0.36^{+.06}_{-.06}$ &  5.0 & $  1.1^{+0.1}_{-0.1}$ & $20.02^{+.13}_{-.11}$ & 6.2 \\ 
\vspace{1mm}
GC8      &    -8.21 &    2.8  &&    $ 0.55^{+ .02}_{- .02}$ &  2.3 & $  38^{+ 2}_{- 2}$ & $15.48^{+.05}_{-.05}$  &&  $-1.34^{+.08}_{-.09}$ &  2.5 & $  2.6^{+0.1}_{-0.1}$ & $13.58^{+.15}_{-.11}$ & 4.0 \\ 
\vspace{1mm}
GC9      &    -7.25 &    7.5  &&    $ 3.25^{+ .16}_{- .20}$ &  6.2 & $  41^{+ 4}_{- 4}$ & $19.74^{+.05}_{-.06}$  &&  $ +0.52^{+.04}_{-.04}$ &  6.0 & $  1.0^{+0.1}_{-0.1}$ & $19.56^{+.09}_{-.07}$ & 6.1\\ 
\vspace{1mm}
GC10    &   -8.49 &    4.0  &&    $ 0.31^{+ .01}_{- .01}$ &  3.2 & $ 139^{+13}_{-12}$ & $14.51^{+.08}_{-.07}$  && $-5.08^{+.26}_{-.54}$ &  3.3 & $  5.5^{+0.1}_{-0.3}$ & $ 8.05^{+.48}_{-.21}$ & 20.5  \\ 
 \hline 

\end{tabular}
\end{minipage}
\end{table*}

\section{Discussion}

In the left panels of Figures 15--17  we show the distribution of King model core,
half-light and tidal radii against projected galactocentric radius
for our sample, compared to the compilations of M31 clusters studied by
\citet{Barmby07}\footnote{We only show those clusters from the
  \citet{Barmby07} sample that are listed as confirmed clusters in the
  Revised Bologna Catalogue V4 \citep{Galleti04}.}  and \citet{strader11}.
 The right
panels show the equivalent data for the Milky Way from
\citet{McLaughlin05} (note, in this case the 3D galactocentric radius
is scaled by $\pi$/4 to make it comparable, in the mean, to the
projected radius used for M31). The M31 sample also includes MGC1, the
most remote globular cluster currently known in M31, which has
well-determined structural parameters from high resolution
ground-based imagery \citep{Mackey10b}. Examining these size estimates
in tandem is relevant since half-light radius is expected to be
relatively insensitive to dynamical evolution, varying little even
through core-collapse \citep[e.g.,][]{gnedin99}.

\begin{figure*}
\begin{center}
\includegraphics[width=17.6cm,angle=0]{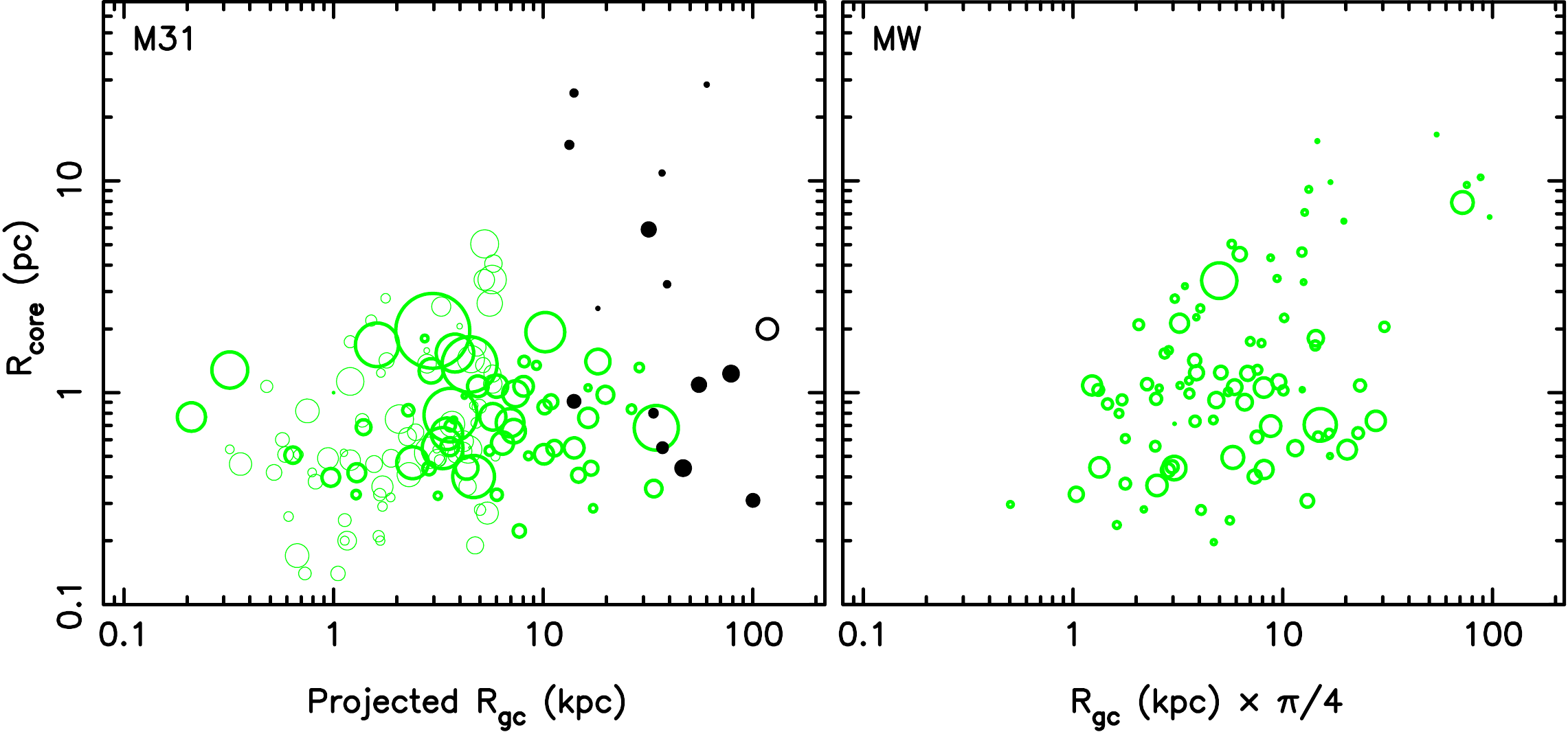}
\caption{
[Left] King model core radius plotted against galactocentric distance (projected)
for M31 globulars.  Solid, black circles are our new data points, 
the open black circle is  MGC1 from \citet{Mackey10b} and
green, open circles are clusters from the compilations of \citet[][bolder symbols]{Barmby07} and 
\citet[][note that we plot their $r_0$ values, and use lighter symbols]{strader11}.  
The area of each circle is proportional to the cluster luminosity.
Unsurprisingly, the four extended clusters stand out as
being very distinct from the population of classical
globular clusters in this figure.
[Right] equivalent figure for the Milky Way GC system, with the galactocentric
radius scaled to an average projected radius.
} 
\end{center}
\end{figure*}

\begin{figure*}
\begin{center}
\includegraphics[width=17.6cm,angle=0]{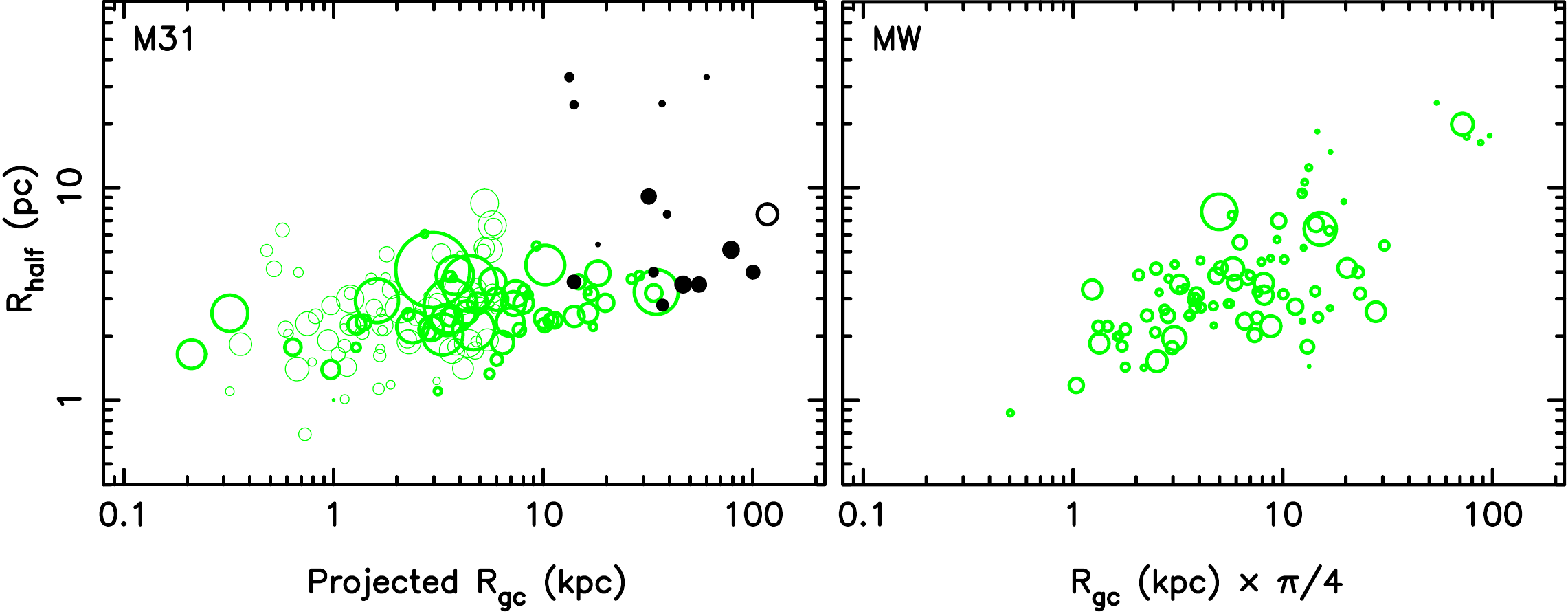}
\caption{
Half light (effective) radius plotted against galactocentric distance (projected)
for M31 and Milky Way globulars.  Symbols 
and panels as for Figure 15. 
}
\end{center}
\end{figure*}

\begin{figure*}
\begin{center}
\includegraphics[width=17.6cm,angle=0]{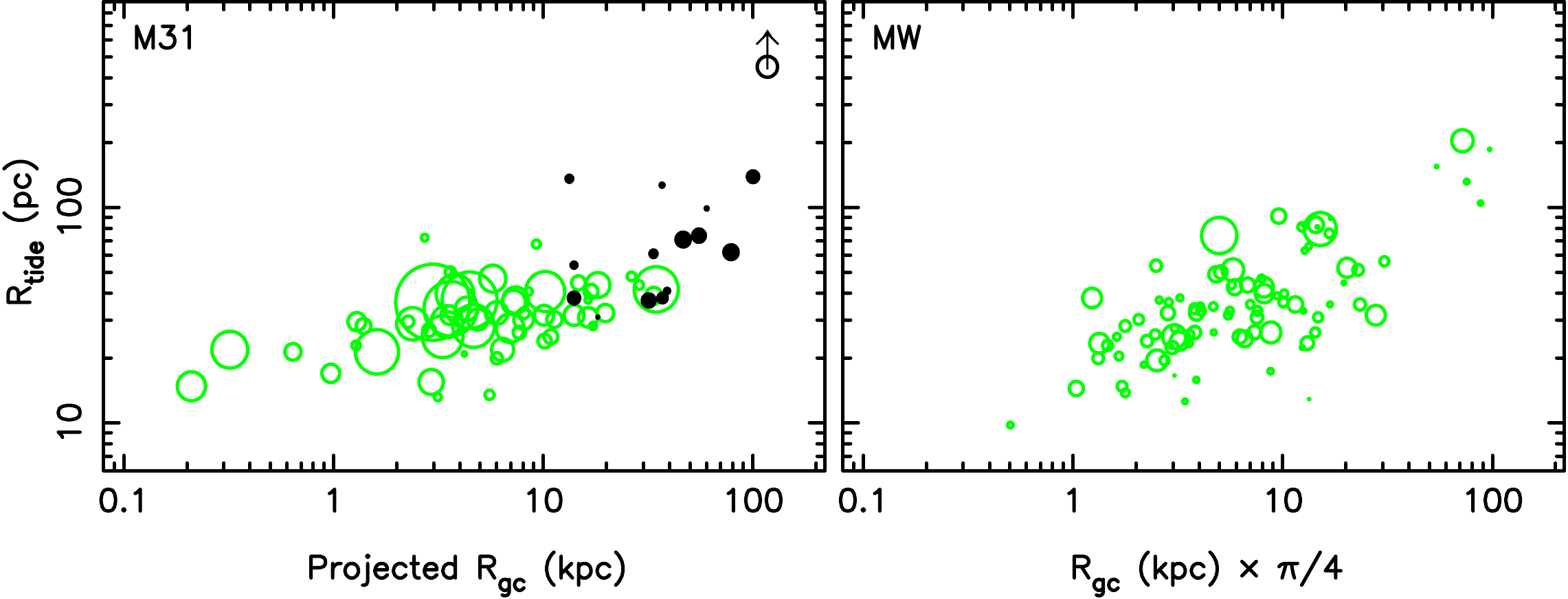}
\caption{
King-model tidal radius plotted against galactocentric distance (projected)
for M31 and Milky Way globulars.  Symbols and panels as for Figure 15
except in this case  the \citet{strader11} clusters are not plotted since
tidal radii were not given in their paper.
} 
\end{center}
\end{figure*}

These samples are not complete, but are sufficiently large to be
broadly representative of the range of cluster properties in each
galaxy.
The extended clusters occupy a unique region of parameter space,
particularly noticeable in Figure 16 where they are quite distinct from the rest of the M31 GC
population.  
Several of the Pal-type clusters in the Milky Way are also comparatively diffuse, but are considerably 
fainter than the M31 extended clusters in our sample.
However it is possible that there is a continuum of sizes
in the extended cluster population, with those objects in the present
sample representing the extreme upper end of the distribution
\citep{Huxor11}.  
Even ignoring the extended clusters, a trend for
increasing half-light radius with galactocentric radius is apparent in
Figure 16 in both systems,  although
it is also clear that in M31,
unlike the Milky Way, at least some clusters beyond $R_{gc}=35$~kpc
are just as compact and luminous as the inner population.  
The presence of such remote compact luminous clusters in M31 has
previously been commented on by \citet{Mackey07} and \citet{Huxor11}
and is confirmed here with our more accurate structural and
photometric measurements.

\subsection{Inner structure of the classical globular clusters}

Four of the classical clusters (GC3, GC6, GC7 and GC9) have convincingly flat
light distributions in their cores.  This is consistent with, the
$\sim$50\% proportion of flat-cored clusters found in the Milky Way
\citep{Noyola06} and the $\sim$60\% proportion in the clusters of MW
satellites \citep{Noyola07}.  Of the others, some show evidence of a
shallow central power-law slope (GC2, GC4, GC8), some a steep central
power-law slope (GC1, GC5, GC10).  As one would expect, all these cases are
better fit by a S\'ersic profile in their inner regions.  
Some caution is appropriate since, as mentioned above, the photometric
error bars in the central $\sim0.1$\,arcsec will certainly not account adequately  for
the graininess due to individual bright stars, but in at least the cases of GC1
and GC10 the central surface brightness is so high that this should be of little concern.

Cuspy cores are usually thought to be the result of post
core-collapse (PCC) evolution.  \citet{Trager95} estimate that 20\% of MW
globulars fall into this category, although criteria for deciding in
individual cases are not well established.
Indeed, recent numerical simulations have shown that PCC
clusters may often be very difficult to distinguish from
pre-core-collapse based on their light profiles
since segregation
of dark remnants in the core leads to a heating and hence spreading
of the visible component
\citep{trenti10}.  A consequence of this is that many more
clusters than previously recognised may be post core-collapse.

Interestingly, our  results suggest
that the spread in the distribution of central surface brightness slopes increases
with increasing radial distance from M31.  Three
of the four GCs at greatest projected distance from M31 have
particularly steep power-law light profiles in their inner regions.
Indeed, the most remote (at least in projection), GC10, has a profile
that is almost a pure power-law over most of the range we probe.
These particularly cuspy clusters contrast markedly with the flat cored
extended clusters (below) which are also found at large galactocentric distance.
We further note that within the Milky Way, the proportion of PCC and other
centrally condensed 
clusters seems to increase significantly toward the centre of the Galaxy
\citep{chernoff89}, possibly due to an increased incidence of tidal
shocking \citep[cf.][]{gnedin99}.  If the steeper profiled outer clusters in M31 
have a similar origin, it would presumably indicate that they
had been stripped from now destroyed satellites where they had
orbited closer to the parent galaxy core.
However, a comparison of the positions of the most steeply cusped
clusters with the regions of known halo substructure from \citet{Mackey10a}
does not show any significant correlation, and hence does not provide
support for such a conclusion.

We should be cautious in comparing these results to those for the
larger sample of inner M31 GCs compiled by \citet{Barmby07} since they
are mostly observed with WFPC2 where the larger pixels make it harder
to quantify inner structure.  However, if we take the S\'ersic index
as one measure of degree of cuspiness, a plot of this against
galactocentric radius (Figure 18, which also includes the extended
clusters) indeed shows a greater spread 
for the outer clusters
(although note only a fraction of
the \citet{Barmby07} sample have S\'ersic fits).

\begin{figure}
\begin{center}
\includegraphics[width=8.4cm,angle=0]{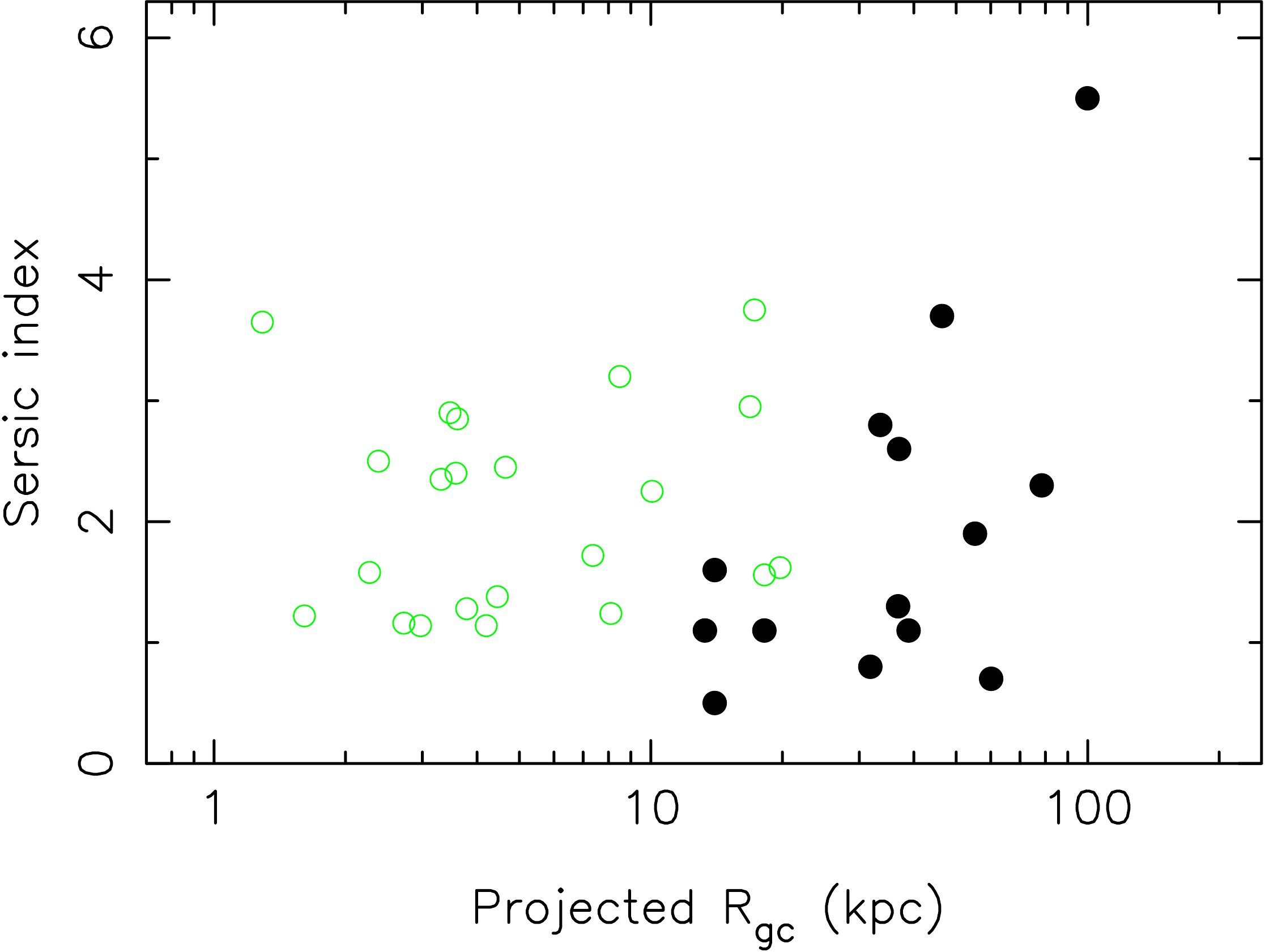}
\caption{
S\'ersic index versus galactocentric radius for M31 clusters.  Filled
symbols are for the ten classical globular clusters reported in this paper, and open
symbols are from \citet{Barmby07}} 
\end{center}
\end{figure}

\subsection{Structure of the extended clusters}

Although the most remote globular clusters in the Milky Way tend to be
faint and diffuse, the discovery of a population of clusters in the
outer parts of M31 that are both brighter and even more extended than
these, was a surprise.  There is, of course, a selection effect
against discovering such clusters at small galactocentric radii, since
they would become increasingly hard to distinguish against a rising background
of crowded field stars.
There is also likely to be a physical selection effect, in the sense that
extended clusters at smaller galactocentric distance would be
vulnerable to complete tidal disruption \citep[e.g.,][]{hurley10},
unless they were embedded in massive dark matter halos, which does not
appear to be the case for EC4 at least \citep{collins09}.

The (circularly averaged) profiles of the four extended clusters studied here are all rather
similar to each other in terms of implied central surface brightness,
core radius and tidal radius.
Although the diffuse nature of these clusters should lead to 
two-body relaxation times which are of order the Hubble time or longer, 
we find they are all well fitted by both King
models with large cores and S\'ersic models with low indices $n\sim1$.
EC3 formally has the smallest tidal radius, although it is in the most
crowded field (in projection it is close to the outer part of M31's
disk), and there is evidence of a lower level excess of cluster stars
beyond the measured tidal radius.  The 2D distribution of stars is
statistically consistent with circular symmetry.

We emphasise that although faint, such central surface brightnesses
are considerably in excess of those of several recently discovered
dwarf galaxies.  The Ursa Major dSph, for example, has a magnitude
similar to the extended clusters studied here, but a half-light radius
ten times larger \citep{Willman05}, whilst other likely dwarf galaxies
are even fainter with significantly lower central surface brightnesses
\citep[e.g.,][]{McConnachie08,Martin08}.  Thus a population of similar
extended clusters in the Milky Way could only have eluded detection if
they were few in number and particularly unfortunately placed for
observation from our location within the Galaxy.

\subsection{Possible asymmetries in EC2 and EC4}

From visual inspection, two of the extended clusters, EC2 and EC4, do show
some indications of 
deviation from circular symmetry (or even point symmetry) in the
distribution of their stars on the sky (see upper right panels of
Figures 2 and 4).  Explaining such deviations by dynamical processes
would be hard, given the age and isolated nature of the clusters, and
so we first ask whether the apparent asymmetries could be simply due to 
statistical chance.

There are a number of difficulties in assessing the significance of
the findings of this sort.
In the first place the
determination of the cluster centroid is affected by counting
statistics 
so it is possible
that the measured centroid is a little out, although, as noted above, 
even for the very sparse EC4, this
uncertainty is unlikely to be more than $\sim2$\,arcsec.
Secondly, the regions of
the image which are masked out around bright stars and galaxies are
not symmetrical around the cluster centre, and their effect depends on
the cluster profile. Finally, and hardest to deal with, is the {\it {\`a}
posteriori} nature of the analysis.  To mitigate this, we try to ask a
fairly general question of the data, namely, if we consider just stars
between the half-light radius and the tidal radius, what is the
maximum ratio of stars in one half of the spatial distribution
relative to the other half, allowing the axis splitting the the sample
across a diameter freedom to be at any angle?  This we compare to the
same statistic determined for each of 10000 simulated clusters, which
are constructed based on the King profile parameters, background star
density and image mask appropriate to the real cluster in question

In the case of EC2 there is an excess of
stars outside the half-light
radius which lie in the positive $x$-direction from the cluster
centre.  Although to the eye this may appear significant, $\sim20$\% of our
simulated clusters showed a similar or greater degree of asymmetry.
In a sample of four extended clusters it is therefore no surprise to
find one like EC2, even if their underlying structure is circularly
symmetric.

The situation with EC4 is potentially even more perplexing, in as much
as it appears to be particularly the horizontal branch stars (and
possibly the red-giant branch stars) outside the half-light radius
which are systematically shifted to higher $x$-values from the overall
centroid.  Once again, the same caveats apply, although in this case
it is even harder to imagine a dynamical explanation for such an
offset.  We do note, however, that the EC4 horizontal branch is
unusual in another way in being particularly broad.  This range of
magnitudes cannot be explained by photometric errors (being the least
crowded and one of the most remote of the clusters, the photometry is
good at these magnitudes), and a line-of-sight depth sufficient to
produce such a spread is highly improbable.  Most likely the
photometric spread of the HB stars is produced by some combination of
the inclusion of several RR Lyrae variables which happen to be caught
close to their maxima, and possibly a small number of contaminating
sources.  To test the significance of such an asymmetry we restrict
our attention to just the 35 horizontal branch stars.  Here we find a
smaller number, 0.2\%, of simulated clusters exhibit an asymmetry of
the same magnitude as the real cluster.  However, again, we must
recognise that we have looked at four extended clusters, and in each
case have split the stars into three or four sub-populations in the
CMD, therefore to find one which shows this level of asymmetry is not
too surprising.

Our conclusion is that the extended clusters are consistent with
being circularly symmetric, and that the marginal evidence for an
asymmetry of the horizontal branch stars with respect to the rest of the cluster in EC4 is not sufficiently
strong with the current data to be regarded as a firm conclusion.

\subsection{Outer structure}

\subsubsection{Tidal radii}

The photometric tidal radii determined from the King profiles generally 
represent the point where the surface brightness finally turns over.
We note that simulations suggest that it is likely
the photometric tidal radius will frequently somewhat overestimate the
true dynamical tidal radius due to the slow evaporation of stars across
the boundary \citep{trenti10}.
On the other hand, a cluster on an orbit which ranges significantly in
galactocentric distance will experience a time-varying tidal field, and
so its present photometric limit might reflect the tidal truncation from an earlier time; 
this effect is considered in more detail below.
Figure 17 shows that tidal radii also
generally have larger values at larger galactocentric radius.
Despite the above caveats, 
such a trend is broadly as expected since the outer clusters sample
the shallower potential gradient at larger distances from
M31. 

GC10 presents a particularly interesting case: it is the
most remote cluster in our sample (in projection) and as remarked previously
has a profile which is almost a pure power-law.
The tidal radius we formally derive is 
$\sim140$\,pc, which is the largest 
of any cluster in our sample,
and since the tidal truncation knee is very poorly defined, 
the photometric tidal radius could easily be even larger.
The outer profile of GC10 is similar to that found using
ground-based data for the cluster MGC1 
which is $\sim 120$\,kpc in projection, and likely at a true distance of
$\sim200$\,kpc, from M31.  In that case the cluster is traced
out to at least 450\,pc and possibly further
\citep{Mackey10b}.

The tidal radius of a globular cluster depends on the potential of the
host galaxy, the potential of the cluster, its orbit around the parent
galaxy, and the orbits of the stars within it \citep[e.g.,][]{read06}.
Here, we parameterise the M31 potential using the model in
\citet{Geehan06},
and the GC potential using a Plummer sphere with half light radius
matched to the GC photometry. We calculate the GC masses from their
$V$-band luminosity, assuming a mass to light ratio of 2
\citep[cf.][]{Pryor93}. We then solve equation 11 of \citet{read06}
for the pericentre $R_{\rm peri}$. There are three other unknowns in this
equation: the apocentre, $R_{\rm apo}$, the 3D distance to the GC,
$x$, and the tidal radius\footnote{We set the internal orbit parameter
  $\alpha=0$ (i.e. radial) since we do not expect significant rotation
  in the GCs \citep[e.g.,][]{Kim08}, and there is no obvious onset of
  tidal tails that might justify using $\alpha=-1$ \citep[i.e.
  retrograde; see][]{read06,Aden09}.}, $R_t$. Since we do not know
these three parameters, we can only calculate a {\it lower bound} on
$R_{\rm peri}$.  For this, we set $R_t$ to its minimum allowed value: the
King tidal radius, assumed to be determined at pericentre, which fixes
$x=R_{\rm peri}$. This leaves only $R_{\rm apo}$ as an unknown.  Here we
use the fact that the GCs are most likely to be at apocentre, since
this is where they spend most of their time on their orbit. We
marginalise over the unknown $R_{\rm apo}$ exploring the range $d <
R_{\rm apo} < 2d$, where $d$ is the projected distance of the GC from
the centre of M31.

Minimum pericentric radii calculated this way are given in Table 2,
 and plotted against present galactocentric distance in Figure 19.
Here we also include clusters from \citet{Barmby07} with projected
galactocentric distance greater than 10\,kpc (below this there is
increasing likelihood that the cluster is a member of the disk
population or has been subject to frequent disk shocking).  The
extended clusters, as expected if they are to avoid disruption,
generally appear to be on low-eccentricity orbits, with pericentric
distances greater than 20\,kpc in three out of four cases.

There is no apparent trend for the most cuspy clusters to have smaller
pericentric radii, which might have been expected if their structure
was due to strong dynamical interactions when at pericentre driving
them to core-collapse.
\citet{gnedin08} find from simulations that in MW-like galaxies all
GCs beyond 10 kpc began life in smaller galaxies, and, in particular,
beyond 50 kpc they are preferentially on radial orbits, and certainly
the outer classical clusters, apart from GC10, apparently have much
more radial orbits than do the extended clusters if we are to believe
the above analysis.
In fact, the \citet{gnedin08} results also suggest that between 10 and
60\,kpc (inner halo) the clusters are primarily from disrupted
satellites, and beyond 60\,kpc they are either still associated with
parent dwarfs or they have been removed via dwarf-dwarf encounters.
The GCs in our study do not seem to be orbiting with known dwarf
galaxies,
although some may be associated with the considerable substructure
which has been revealed by recent M31 surveys
\citep[e.g.,][]{chapman08,Mackey10a}.

\subsubsection{Tidal evaporation}

We can use the spatial distribution of likely cluster stars to search
for possible tidal distortion and extra-tidal stars indicating
significant ongoing evaporation.  The distributions are shown in the
upper-right panels of Figures 1--14.  In several instances there is a
suggestion of such distortion.  Perhaps the best case is that of GC4,
which has an excess of stars beyond the photometric tidal radius out to about
300\,pc, particularly in the east-west direction.  This cluster has also been studied by \citet{Federici07},
who came to a very similar conclusion.  It is not obvious why GC4
should exhibit a significant extratidal component, given that it is very remote
from M31 ($\sim$50~kpc in projection) along the major axis.  

The other good case is GC5, at an even greater distance from M31,
which exhibits a similar, albeit less pronounced, excess.
An apparently significant excess of extra-tidal stars is also seen in
EC3.  However, since the background to this field is a relatively
crowded region of the outer disk of M31, it seems more likely that the
excess is due to a gradient in the density of contaminating stars.
Several other clusters, e.g. GC8 and GC9, show marginal evidence for a
small excess beyond the formal King-model tidal radius.

On the other hand \citet{McLaughlin05,McLaughlin08} have shown that many globular clusters 
are better modelled 
by \citet{wilson75} profiles, which have greater extension in the outer parts than \citet{King62} or \citet{king66}
models.  
Whilst the theoretical basis of Wilson models is somewhat {\it ad-hoc}, it may indicate that
clusters such as GC4 and GC5 are not overflowing their tidal radii.
However, for GC4, in particular, the roughly elliptical distribution of the stars
beyond the King tidal radius does remain suggestive of a tidal process.

\subsubsection{Are any clusters embedded in dark matter halos?}

Although measurements of mass-to-light ratios in GCs generally do not
find evidence of their being embedded in extended dark matter halos,
it has been suggested that clusters may have formed that way
\citep{peebles84} but with the dark halos being
largely tidally stripped during their subsequent journeys through
the parent galaxy potential.

\citet{conroy11} have recently argued that for GCs at large
galactocentric distance, which may therefore have suffered little
tidal stripping in their lifetimes, the existence of an extended dark
matter halo should manifest itself in a shallow gradient of outer
stellar density.  They specifically considered MGC1, at a distance of
$\sim200$\,kpc from M31 \citep{Mackey10b}, and showed that the slope of
projected stellar surface density declines more steeply than
$R^{-5/2}$ for radii larger than $\sim20$\,pc.  This compares to an
expected slope at least as shallow as $R^{-3/2}$ for high $M/L$ models
in the same regime of radius, and so they conclude this cluster is not
dark matter dominated.  Although our clusters appear not to be at such
great galactocentric radius as MGC1, we note that the GCs at largest
projected distance (GCs 1, 4, 5 and 10) all have outer slopes at least
as steep as $R^{-5/2}$, and so we similarly conclude that they show no
evidence of residing in their own dark matter halos.

\begin{figure}
\begin{center}
\includegraphics[width=8.4cm,angle=0]{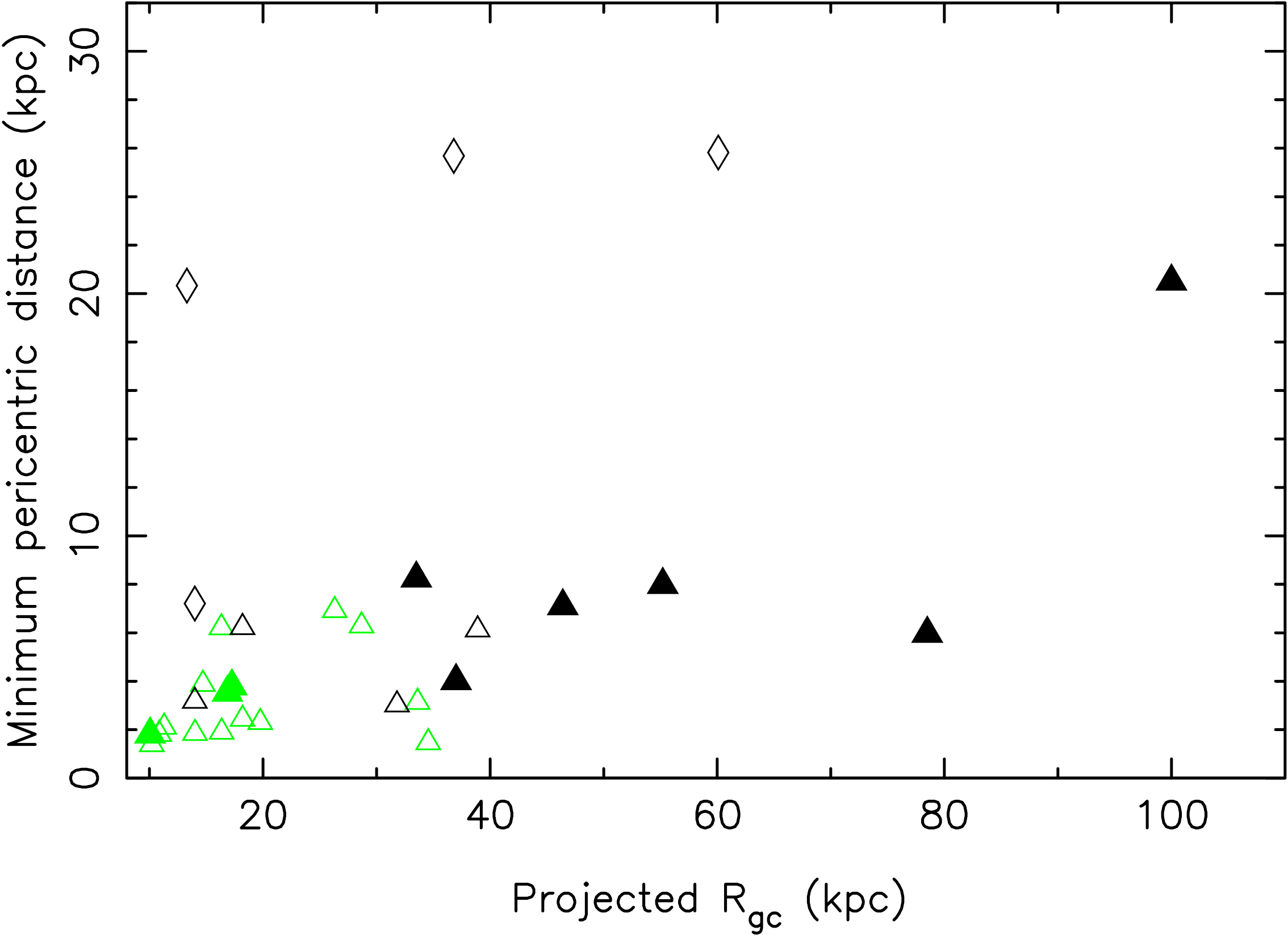}
\caption{
Inferred minimum pericentric distance (see text) versus its 
current galactocentric distance, for all  M31 globular (triangles) and 
extended (diamonds) clusters with projected $R_{gc}>10$~kpc.  
Black symbols are our new sample of clusters reported in this paper, and green are those from
\citet{Barmby07}.
Filled symbols are cuspy clusters with S\'ersic index
$n>2$.
} 
\end{center}
\end{figure}

%\end{document}

\section{Conclusions}

We must be aware that neither our sample, nor the
sample of M31 GCs studied by other groups, are well
defined in any statistical sense.  Generally they
are biased to brighter magnitudes, for example.
Nonetheless, although selection criteria are not
homogeneous, the total sample is probably reasonably
representative of at least the brighter population
of normal GCs.  In addition to that, a number of exceptional
clusters, such as the extended clusters reported here, 
allow us to study properties of relatively extreme
objects from the overall population.

The findings of our study are:
\begin{itemize}
\item The extended clusters are well fit by both King and S\'ersic profiles with
large core radii, around 10--30\,pc, confirming they are very
distinct from the bulk of classical GCs. 
Three of the four have large King-model tidal radii of
about 100\,pc,
which is consistent with their
being on relatively circular orbits where they experience
relatively little tidal stress.
\item The measured photometric tidal radii of the classical globulars are also typically
50--100\,pc.  Together with the ECs, the  apparent  trend of increasing tidal radius with increasing
with galactocentric radius is as expected if the photometric measures are providing
a reasonable indication of dynamical tidal truncation.  
In at least two cases, GC4 and GC5,  there
is evidence of a halo of stars beyond the King-model tidal radius, although 
as discussed in the text, for individual cases it is not necessarily the case
that this photometric model parameter corresponds well to the dynamical tidal radius.
\item Overall, the properties of our outer sample extend the
trend of larger half-light and core radii with increasing galactocentric
distance seen in samples of inner GCs in M31, and also in
the MW GC system.  But they also reinforce the finding that in
M31, the range of cluster properties (size and luminosity) is
much greater in the outer halo of M31 than it is in the Milky Way.
\item In particular, we confirm that M31 possesses
  a  number of compact, luminous clusters at large galactocentric radii 
  that 
  have no counterparts in the MW.
\item About half of the GCs are cuspy in their central
regions, and hence not well-fit by King profiles.  In particular,
there seems to be a trend for some of the most remote clusters to have
the steepest central power-law slopes.
Thus the spread in the cuspiness for the whole population of 
clusters seems to increase at large galactocentric radius.
Steep central surface brightness gradients may indicate these clusters
have been through core-collapse.  

\end{itemize}

\section*{Acknowledgments}

NRT acknowledges a STFC Senior Research Fellowship.

AMNF and ADM acknowledge support from a Marie Curie Excellence Grant
from the European Commission under contract MCEXT-CT-2005-025869.

ADM acknowledges an Australian Research Fellowship.

GFL thanks the Australian Research Council for support through his
Future Fellowship (FT100100268) and Discovery Project (DP110100678).

Based on observations made with the NASA/ESA {\em Hubble Space
Telescope}, obtained at the Space Telescope Science Institute (STScI),
which is operated by the Association of Universities for Research
in Astronomy, Inc., under NASA contract NAS 5-26555.  These observations
are associated with program GO-10394.

\bsp

\label{lastpage}


\begin{thebibliography}{99}


\bibitem[\protect\citeauthoryear{Ad{\'e}n et 
al.}{2009}]{Aden09} Ad{\'e}n D., Wilkinson M.~I., Read J.~I., 
Feltzing S., Koch A., Gilmore G.~F., Grebel E.~K., Lundstr{\"o}m I., 2009, 
ApJ, 706, L150

\bibitem[\protect\citeauthoryear{Barmby, Holland, \& 
Huchra}{2002}]{Barmby02} Barmby P., Holland S., Huchra J.~P., 
2002, AJ, 123, 1937 

\bibitem[\protect\citeauthoryear{Barmby et al.}{2007}]{Barmby07} 
Barmby P., McLaughlin D.~E., Harris W.~E., Harris G.~L.~H., Forbes D.~A., 
2007, AJ, 133, 2764 

\bibitem[\protect\citeauthoryear{Brasseur et al.}{2011}]{brasseur11} Brasseur C.~M., Martin N.~F., Macci{\`o} 
A.~V., Rix H.-W., Kang X., 2011, ApJ, 743, 179 


\bibitem[\protect\citeauthoryear{Chapman et 
al.}{2005}]{Chapman05} Chapman S.~C., Ibata R., Lewis G.~F., 
Ferguson A.~M.~N., Irwin M., McConnachie A., Tanvir N., 2005, ApJ, 632, L87 

\bibitem[\protect\citeauthoryear{Chapman et al.}{2008}]
{chapman08} Chapman S.~C., et al., 2008, MNRAS, 390, 
1437 

\bibitem[\protect\citeauthoryear{Chernoff et al.}{1989}]
{chernoff89} Chernoff D.~F. \& Djorgovski, S, 1989, ApJ, 339, 904 

\bibitem[\protect\citeauthoryear{Cohen et al.}{2010}]{cohen10} 
Cohen J.~G., Kirby E.~N., Simon J.~D. \& Geha M., 2010, ApJ, 725, 288 

\bibitem[\protect\citeauthoryear{Collins et al.}{2009}]
{collins09} Collins M.~L.~M., et al., 2009, MNRAS, 396, 1619 

\bibitem[\protect\citeauthoryear{Collins et 
al.}{2011}]{collins11} Collins M.~L.~M., et al., 2011, MNRAS, 
417, 1170 

\bibitem[\protect\citeauthoryear{Conroy, Loeb, \& Spergel}{2011}]
{conroy11} Conroy C., Loeb A., Spergel D.~N., 2011, ApJ, 741, 72 


\bibitem[\protect\citeauthoryear{Federici et 
al.}{2007}]{Federici07} Federici L., Bellazzini M., Galleti S., 
Fusi Pecci F., Buzzoni A., Parmeggiani G., 2007, A\&A, 473, 
429

\bibitem[\protect\citeauthoryear{Ferguson et 
al.}{2002}]{Ferguson02} Ferguson A.~M.~N., Irwin M.~J., Ibata 
R.~A., Lewis G.~F., Tanvir N.~R., 2002, AJ, 124, 1452 

\bibitem[\protect\citeauthoryear{Galleti et al.}{2004}]{Galleti04}
Galleti S., Federici L., Bellazzini M., Fusi Pecci F., Macrina S.,
2004, A\&A, 416, 917

\bibitem[\protect\citeauthoryear{Geehan et al.}{2006}]{Geehan06} 
Geehan J.~J., Fardal M.~A., Babul A., Guhathakurta P., 2006, MNRAS, 366, 
996 

\bibitem[\protect\citeauthoryear{Gilbert et 
al.}{2006}]{Gilbert06} Gilbert K.~M., et al., 2006, ApJ, 652, 
1188 

\bibitem[\protect\citeauthoryear{Gnedin, Lee, 
\& Ostriker}{1999}]{gnedin99} Gnedin O.~Y., Lee H.~M., Ostriker J.~P., 1999, ApJ, 522, 935 

\bibitem[\protect\citeauthoryear{Gnedin 
\& Prieto}{2008}]{gnedin08} Gnedin O.~Y., Prieto J.~L., 2008, IAUS, 246, 403 

\bibitem[\protect\citeauthoryear{Grillmair et 
al.}{1996}]{Grillmair96} Grillmair C.~J., Ajhar E.~A., Faber S.~M., 
Baum W.~A., Holtzman J.~A., Lauer T.~R., Lynds C.~R., O'Neil E.~J., Jr., 
1996, AJ, 111, 2293 


\bibitem[\protect\citeauthoryear{Holland}{1998}]{Holland98} 
Holland S., 1998, PASP, 110, 759 

\bibitem[\protect\citeauthoryear{Hurley \& Mackey}{2010}]{hurley10} 
Hurley J.~R., Mackey A.~D., 2010, MNRAS, 408, 2353 

\bibitem[\protect\citeauthoryear{Huxor et al.}{2004}]{Huxor04} 
Huxor A., Tanvir N.~R., Irwin M., Ferguson A., Ibata R., Lewis G., Bridges 
T., 2004, in
Satellites and Tidal Streams, Astronomical Society of the Pacific. Eds 
F.~Prada, D.~Martinez~Delgado, and T.~J.~Mahoney, 327, 118 

\bibitem[\protect\citeauthoryear{Huxor et al.}{2005}]{Huxor05} 
Huxor A.~P., Tanvir N.~R., Irwin M.~J., Ibata R., Collett J.~L., Ferguson 
A.~M.~N., Bridges T., Lewis G.~F., 2005, MNRAS, 360, 1007 

\bibitem[\protect\citeauthoryear{Huxor et al.}{2011}]{Huxor11} Huxor
  A.~P., et al., 2011, MNRAS, 414, 770

\bibitem[\protect\citeauthoryear{Huxor et al.}{2008}]{Huxor08} 
Huxor A.~P., Tanvir N.~R., Ferguson A.~M.~N., Irwin M.~J., Ibata R., 
Bridges T., Lewis G.~F., 2008, MNRAS, 385, 1989 

\bibitem[\protect\citeauthoryear{Ibata et al.}{2001}]{Ibata01} 
Ibata R., Irwin M., Lewis G., Ferguson A.~M.~N., Tanvir N., 2001, Nature, 
412, 49 


\bibitem[\protect\citeauthoryear{Ibata et al.}{2005}]{Ibata05} 
Ibata R., Chapman S., Ferguson A.~M.~N., Lewis G., Irwin M., Tanvir N., 
2005, ApJ, 634, 287 

\bibitem[\protect\citeauthoryear{Ibata et al.}{2007}]{Ibata07} 
Ibata R., Martin N.~F., Irwin M., Chapman S., Ferguson A.~M.~N., Lewis 
G.~F., McConnachie A.~W., 2007, ApJ, 671, 1591 

\bibitem[\protect\citeauthoryear{Irwin et al.}{2005}]{Irwin05} 
Irwin M.~J., Ferguson A.~M.~N., Ibata R.~A., Lewis G.~F., Tanvir N.~R., 
2005, ApJ, 628, L105 

\bibitem[\protect\citeauthoryear{Irwin et al.}{2008}]{Irwin08} 
Irwin M.~J., Ferguson A.~M.~N., Huxor A.~P., Tanvir N.~R., Ibata R.~A., 
Lewis G.~F., 2008, ApJ, 676, L17 

\bibitem[\protect\citeauthoryear{Kim et al.}{2008}]{Kim08} 
Kim E., Yoon I., Lee H.~M., Spurzem R., 2008, MNRAS, 383, 2 

\bibitem[\protect\citeauthoryear{King}{1962}]{King62} King I., 
1962, AJ, 67, 471 

\bibitem[\protect\citeauthoryear{King}{1966}]{king66} King 
I.~R., 1966, AJ, 71, 64 

\bibitem[\protect\citeauthoryear{Mackey \& 
Gilmore}{2003}]{Mackey03a} Mackey A.~D., Gilmore G.~F., 2003, 
MNRAS, 338, 85 

\bibitem[\protect\citeauthoryear{Mackey \& 
Gilmore}{2003}]{Mackey03b} Mackey A.~D., Gilmore G.~F., 2003, 
MNRAS, 340, 175 

\bibitem[\protect\citeauthoryear{Mackey \& Gilmore}{2004}]{Mackey04}
Mackey A.~D., Gilmore G.~F., 2004, MNRAS, 355, 504

\bibitem[\protect\citeauthoryear{Mackey \& van den Bergh}{2005}]
{mackey05} Mackey A.~D., van den Bergh S., 2005, MNRAS, 360, 631 

\bibitem[\protect\citeauthoryear{Mackey et al.}{2006}]{Mackey06} 
Mackey A.~D., et al., 2006, ApJ, 653, L105 

\bibitem[\protect\citeauthoryear{Mackey et al.}{2007}]{Mackey07} 
Mackey A.~D., et al., 2007, ApJ, 655, L85 

\bibitem[\protect\citeauthoryear{Mackey et al.}{2010a}]{Mackey10a} 
Mackey A.~D., et al., 2010a, ApJ, 717, L11

\bibitem[\protect\citeauthoryear{Mackey et al.}{2010b}]{Mackey10b} 
Mackey A.~D., et al., 2010b, MNRAS, 401, 533 

\bibitem[\protect\citeauthoryear{Martin et al.}{2006}]{Martin06} 
Martin N.~F., Ibata R.~A., Irwin M.~J., Chapman S., Lewis G.~F., Ferguson 
A.~M.~N., Tanvir N., McConnachie A.~W., 2006, MNRAS, 371, 1983 

\bibitem[\protect\citeauthoryear{Martin, de Jong, 
\& Rix}{2008}]{Martin08} Martin N.~F., de Jong J.~T.~A., Rix H.-W., 2008, ApJ, 684, 1075 

\bibitem[\protect\citeauthoryear{Martin et al.}{2009}]{Martin09} Martin N.~F., et al., 2009, 
ApJ, 705, 758 

\bibitem[\protect\citeauthoryear{McConnachie et 
al.}{2005}]{mcconnachie05} McConnachie A.~W., Irwin M.~J., Ferguson 
A.~M.~N., Ibata R.~A., Lewis G.~F. \& Tanvir N., 2005, MNRAS, 356, 979 

\bibitem[\protect\citeauthoryear{McConnachie \& Irwin}{2006}]{McConnachie06} McConnachie A.~W. \& Irwin, M.~J., 2006, MNRAS, 365, 
1263 

\bibitem[\protect\citeauthoryear{McConnachie et 
al.}{2008}]{McConnachie08} McConnachie A.~W., et al., 2008, ApJ, 688, 
1009 

\bibitem[\protect\citeauthoryear{McConnachie et al.}{2009}]{McConnachie09} 
McConnachie A.~W., et al., 2009, Nature, 461, 66 


\bibitem[\protect\citeauthoryear{McLaughlin \& van der 
Marel}{2005}]{McLaughlin05} McLaughlin D.~E., van der Marel R.~P., 
2005, ApJS, 161, 304 

\bibitem[\protect\citeauthoryear{McLaughlin et 
al.}{2008}]{McLaughlin08} McLaughlin D.~E., Barmby P., Harris W.~E., 
Forbes D.~A., Harris G.~L.~H., 2008, MNRAS, 384, 563 

\bibitem[\protect\citeauthoryear{Noyola \& Gebhardt}{2007}]{Noyola07}
Noyola E., Gebhardt K., 2007, AJ, 134, 912

\bibitem[\protect\citeauthoryear{Noyola \& Gebhardt}{2006}]{Noyola06} Noyola E., Gebhardt K., 2006, AJ, 
132, 447 

\bibitem[\protect\citeauthoryear{Peebles}{1984}]{peebles84} 
Peebles P.~J.~E., 1984, ApJ, 277, 470 

\bibitem[\protect\citeauthoryear{Pryor \& Meylan}{1993}]{Pryor93} Pryor C., Meylan G., 1993, ASPC, 50, 357 

\bibitem[\protect\citeauthoryear{Read et al.}{2006}]{read06} 
Read J.~I., Wilkinson M.~I., Evans N.~W., Gilmore G., Kleyna J.~T., 2006, 
MNRAS, 366, 429 

\bibitem[\protect\citeauthoryear{Richardson et
    al.}{2009}]{Richardson09} Richardson J.~C., et al., 2009, MNRAS,
  396, 1842

\bibitem[\protect\citeauthoryear{Richardson et
    al.}{2011}]{Richardson11} Richardson J.~C., et al., 2011, ApJ,
  732, 76

\bibitem[\protect\citeauthoryear{S\'ersic}{1968}]{Sersic68} S\'ersic 
J.~L., 1968, Atlas de Galaxias Australes (Cordoba: Observatoria Astronomico).  

\bibitem[\protect\citeauthoryear{Strader, Caldwell,  \& Seth}{2011}]{strader11} 
Strader J., Caldwell N., Seth A.~C., 2011, AJ, 142, 8 

\bibitem[\protect\citeauthoryear{Trager, King, \& 
Djorgovski}{1995}]{Trager95} Trager S.~C., King I.~R., Djorgovski S., 1995, AJ, 109, 1912 

\bibitem[\protect\citeauthoryear{Trenti et al.}{2010}]{trenti10}
Trenti, M., Vesperini, E. \& Pasquato, M., 2010, ApJ, 708, 1958

\bibitem[\protect\citeauthoryear{van den Bergh \& Mackey}{2004}]{vdB04} 
van den Bergh S., Mackey A.~D., 2004, MNRAS, 354, 713

\bibitem[\protect\citeauthoryear{Willman et al.}{2005}]{Willman05}
Willman B., et al., 2005, ApJ, 626, L85

\bibitem[\protect\citeauthoryear{Wilson}{1975}]{wilson75} Wilson 
C.~P., 1975, AJ, 80, 175 

\bibitem[\protect\citeauthoryear{Zinn}{1993}]{Zinn93} Zinn R., 1993,
in Smith G.H., Brodie, J.P., eds., ASP Conf. Ser. 48: The Globular
Cluster-Galaxy Connection. Astron. Soc. Pac., San Francisco, p. 38

\bibitem[\protect\citeauthoryear{Zucker et al.}{2004}]{Zucker04} 
Zucker D.~B., et al., 2004, ApJ, 612, L121 


\end{thebibliography}
\end{document}